\documentclass[11pt]{article}

\usepackage[margin=1in]{geometry}
\usepackage{indentfirst}
\usepackage{graphicx}
\usepackage{amsmath}
\usepackage{setspace}
\usepackage{hyperref}
\usepackage{caption}

\hypersetup{
  colorlinks=true,
  linkcolor=black,
  citecolor=black,
  urlcolor=black
}

\title{Atomically Precise Electron Beam Sculpting of Bilayer h-BN: The Role of Crystallographic Orientation and Milling Strategy}

\author{%
\texorpdfstring{%
\textit{Ondrej Dyck, Andrew R. Lupini, Ivan Vlassiouk, Matthew Brahlek, Rob Moore, Stephen Jesse}\\[4pt]
\textit{\small Center for Nanophase Materials Sciences, Oak Ridge National Laboratory, Oak Ridge, TN}\\
\textit{\small Correspondence: \href{mailto:dyckoe@ornl.gov}{dyckoe@ornl.gov}}%
}{%
Ondrej Dyck, Andrew R. Lupini, Ivan Vlassiouk, Matthew Brahlek, Rob Moore, Stephen Jesse%
}%
}

\date{}

\begin{document}
\maketitle

\begingroup
\renewcommand{\thefootnote}{}
\renewcommand{\theHfootnote}{cc0}
\footnotetext{\includegraphics[height=0.9em]{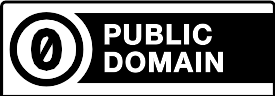}\,Released into the public domain under CC0.}
\addtocounter{footnote}{-1}
\endgroup


\begin{abstract}
Achieving atomic precision in top-down manufacturing remains a fundamental challenge nanofabrication technology. Here, the focused electron beam of a scanning transmission electron microscope is used to demonstrate atomically precise sculpting of hexagonal boron nitride (h-BN) bilayers, achieving nanoribbons as narrow as 6 \AA{} with atomically smooth edges. The key to this precision lies in understanding how the underlying atomic structure, particularly in twisted bilayer systems, influences the milling process. High-angle annular dark-field imaging combined with multislice simulations reveals distinct intensity signatures that allow identification of different stacking arrangements within moir\'{e} patterns. Mathematical analysis of moir\'{e} lattices provides a predictive framework for determining optimal cutting directions, with cuts along armchair directions yielding superior edge quality compared to zigzag orientations. Surprisingly, a sequential milling approach, where a small electron beam subscan area is translated during the process, produces significantly better results than parallel milling of the entire target region. To understand these differences we implemented a stochastic milling model that reveals that sequential milling minimizes unwanted exposure to surrounding material through beam tail effects. These findings establish a framework for achieving atomic precision in electron beam sculpting of two-dimensional materials and provide fundamental insights applicable to the broader challenge of top-down nanofabrication.
\end{abstract}

\clearpage
\begingroup
\begin{singlespace}
\setcounter{tocdepth}{2}
\tableofcontents
\end{singlespace}
\endgroup
\clearpage


\section{Introduction}
\label{sec:introduction}

\subsection{Background}
\label{sec:intro_background}

The quest for atomic precision in manufacturing has driven technological advancement from the development of transistors to modern quantum devices. While bottom-up approaches such as chemical synthesis and molecular assembly have achieved remarkable atomic control~\cite{groning2018,rizzo2020,marangoni2016,mccurdy2021}, top-down manufacturing---the subtractive process of removing material to create desired structures---has lagged significantly behind in precision. Traditional lithographic techniques, even at their most advanced, operate with feature sizes orders of magnitude larger than individual atoms. This limitation represents a fundamental bottleneck in the fabrication of next-generation devices where atomic-scale control over structure and interfaces is paramount.

Two-dimensional (2D) materials present a unique opportunity to bridge this precision gap. Their atomically thin nature means that even modest improvements in fabrication control can yield structures with true atomic precision. Among 2D materials, hexagonal boron nitride (h-BN) has emerged as particularly important due to its wide bandgap, excellent thermal stability, and its role as an ideal substrate and encapsulating layer for other 2D materials~\cite{dean2010}. The ability to pattern h-BN with atomic precision could support the creation of quantum confinement structures, controlled edge states, and precisely defined heterointerfaces. This capability would pair excellently with atomic precision doping~\cite{dyck2023} or edge passivation~\cite{elibol2023}.

Electron beam lithography and sculpting represent one of the most promising avenues toward atomic precision in top-down processing~\cite{vandorp2005,vandorp2012,sang2018,lin2016}. The focused electron beam in a scanning transmission electron microscope (STEM) can be positioned with sub-angstrom accuracy~\cite{roccapriore2025} and can selectively remove atoms through knock-on damage and sputtering processes. However, achieving controlled, reproducible atomic precision with electron beam sculpting has proven challenging due to the stochastic nature of the atomic ejection process and the complex relationship between beam parameters, material structure, and the resulting cuts.

Recent advances in understanding twisted bilayer systems have revealed that the local atomic arrangement---particularly the stacking configuration and moir\'{e} patterns that emerge from layer misalignment---can dramatically influence material properties and processing outcomes~\cite{carr2020}. These insights suggest that the key to achieving atomic precision in electron beam sculpting may lie not just in optimizing beam parameters, but in understanding and exploiting the underlying atomic structure of the target material.

\subsection{The Monolayer Challenge and Bilayer Strategy}
\label{sec:intro_monolayer}

\begin{figure}
\centering
  \includegraphics[scale=0.8]{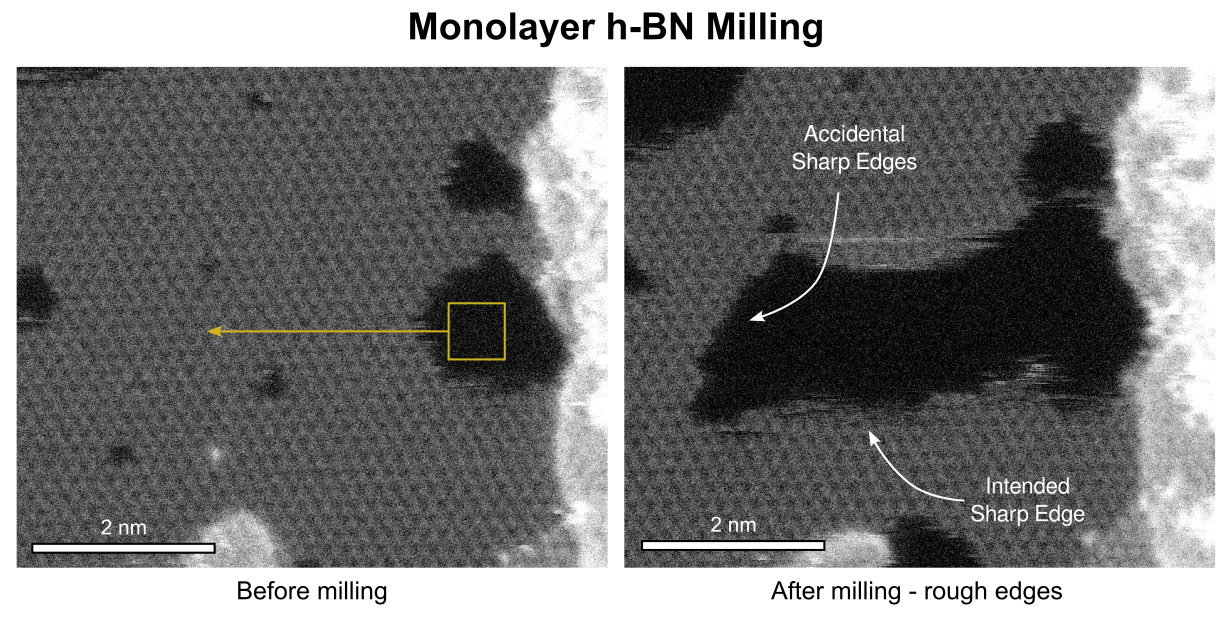}
  \caption{Electron beam milling of monolayer h-BN. (a) HAADF image before milling. The boxed region shows the subscan area with the arrow indicating the direction of movement of the subscan box. (b) After milling, the top and bottom edges show irregular atomic structure even though this is the crystallographically preferred milling direction. The unsatisfactory result motivated the investigation of bilayer systems, where interlayer coupling might provide additional structural constraints.}
  \label{fig1}
\end{figure}

Initial attempts to achieve atomic precision through electron beam milling of monolayer h-BN proved unsatisfactory. Figure~\ref{fig1} shows representative results: a subscan box was defined as indicated and dragged across the sample to eject atoms from the h-BN. The top and bottom edges exhibit irregular atomic-scale structure. We can see that this crystallographic direction (Zig-Zag (ZZ)) is the preferred direction. While spontaneous etching of h-BN under electron beam exposure occasionally produces atomically sharp edges, as also exhibited here, intentionally reproducing these results through directed milling remained elusive. The stochastic nature of atomic ejection, combined with the lack of structural constraints at monolayer edges, appeared to preclude reliable atomic-scale control.

This failure motivated the investigation of bilayer systems, where interlayer coupling might stabilize edge configurations during the milling process. Initial experiments with twisted bilayer h-BN revealed that atomically sharp edges could indeed be achieved---a striking contrast to the monolayer results. However, interpreting these results and identifying optimal milling conditions required addressing several fundamental questions. First, how can we reliably identify different stacking arrangements within twisted bilayer systems from HAADF image contrast? Second, what is the mathematical relationship between crystallographic directions in the individual layers and the emergent moir\'{e} pattern? Third, beyond crystallographic orientation, what role do milling strategies---such as sequential versus parallel approaches---play in achieving atomic precision? This last question was not anticipated at the outset but emerged as a critical factor in understanding the experimental results.

To address these questions, we combine advanced electron microscopy imaging, mathematical analysis of moir\'{e} lattice structures, and systematic investigation of electron beam milling strategies to achieve atomically precise sculpting of h-BN bilayers. Through careful analysis of imaging contrast and comparison with multislice simulations, distinct structural signatures are identified that allow precise determination of local stacking arrangements. Mathematical treatment of moir\'{e} patterns provides a rigorous framework for understanding how crystallographic directions in the substrate relate to directions in the moir\'{e} lattice. Finally, systematic comparison of different milling strategies reveals fundamental principles that enable the creation of nanoribbons as narrow as 6 \AA{} with atomically smooth edges. These results establish a framework for atomic precision in top-down nanofabrication that leverages the structural constraints provided by bilayer systems, with insights applicable to the broader challenge of controllable atomic manipulation.

\section{Twisted Bilayer h-BN}
\label{sec:twisted_bilayer}
\subsection{Interpreting HAADF Image Contrast of Twisted Bilayer h-BN}
\label{sec:haadf_contrast}
In order to properly interpret high angle annular dark-field (HAADF) images of h-BN, especially images of bilayers, it is instructive to carefully examine the underlying structure, types of stacking, and perform simulations that capture the expected contrast. In particular, the contrast of HAADF is proportional to the atomic number of the atom under examination, which, to a first approximation, is proportional to z$^2$ (z being the atomic number). Since h-BN is composed of B (z = 5) and N (z = 7), we can directly observe the difference between the two types of atoms in the image contrast, N being brighter than B. However, with stacked and especially twisted bilayers, the interpretation of image contrast is less straightforward because of the different stacking possibilities. In this section, we examine the expected contrast from stacked h-BN with a particular focus on the twisted bilayer case.

We have already mentioned that HAADF contrast is proportional to the atomic number, with heavier atoms exhibiting stronger scattering and appearing brighter. An additional source of contrast is thickness contrast. Atoms stacked on top of each other will appear brighter than single atoms of the same type. Assuming additive contributions to the total intensity, we expect that two B atoms stacked on top of each other will be twice as bright as a single B atom. Thus, a twisted bilayer h-BN sample will display a variety of different intensities depending on the atomic species and stacking order. In this section, our aim is to clarify the sources of various intensities observed in our sample.

\begin{figure}
\centering
  \includegraphics[width=\linewidth,height=0.72\textheight,keepaspectratio]{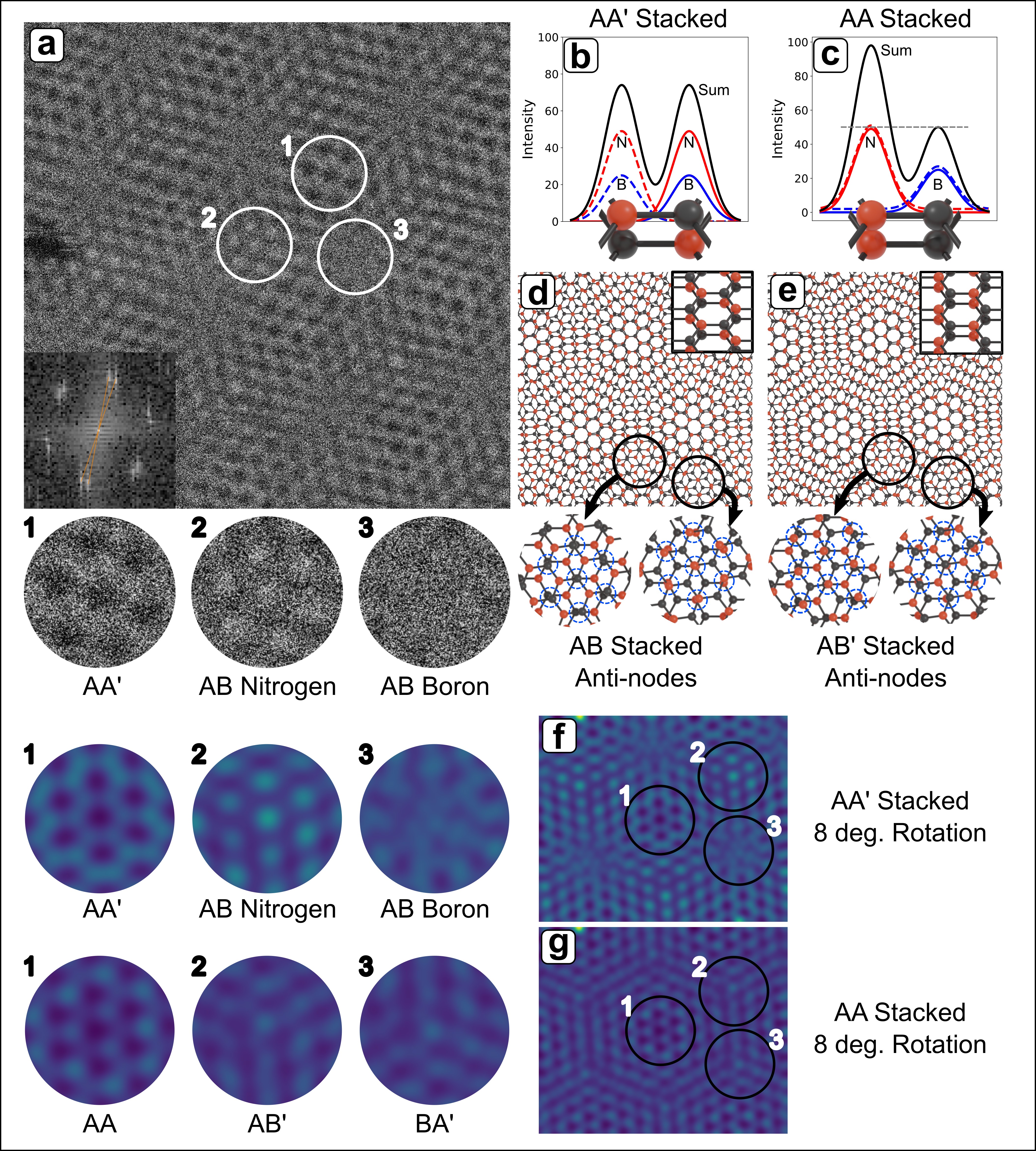}
  \caption{Analysis of twisted bilayer h-BN image intensity features. (a) HAADF image of twisted bilayer h-BN. The fast Fourier transform is overlaid, from which we can measure the twist angle ($\sim$ 8 deg.). The nodes marked 1-3 are magnified below. (b) Schematic illustration the expected intensity of an AA$'$ stacked region, assuming a Gaussian shape and z\textsuperscript{2} intensity. B intensity is shown in blue; N intensity is shown in red; the sum of two layers is shown in black. (c) Schematic illustration of the expected intensity of an AA stacked region. Dotted horizontal line is a guide for the eye, illustrating that two B atoms are approximately the same intensity as a single N atom. Note, (b) and (c) are not simulated intensities. They are intended as a schematic representation only. (d) Model of an AA$'$ stacked bilayer with an 8 deg. rotation. One set of nodes maintain AA$'$ stacking, while the other two nodes exhibit AB stacking; B above B in one node, and N above N in the other, highlighted by the blue circles. (e) Model of an AA stacked bilayer with an 8 deg. rotation. One set of nodes maintain AA stacking, while the nodes exhibit AB' stacking; B above N in one anti-node and N above B in the other anti-node, highlighted by the blue circles. (f) Multislice simulation of AA$'$ stacked h-BN with an 8 deg. rotation. Nodes marked 1-3 are shown magnified to the left for comparison with those from the experimental image (a). (g) Multislice simulation of AA stacked h-BN with an 8 deg. rotation. Nodes marked 1-3 are shown to the left.} 
  \label{fig2}
\end{figure}

Figure~\ref{fig2}(a) shows a typical HAADF image of a twisted bilayer with a twist angle of $\sim$8 degrees. The inset shows the fast Fourier transform (FFT) where we can identify the double hexagonal pattern corresponding to the two hexagonal h-BN layers. The rotation angle between the two layers was measured by overlaying two lines on the FFT, crossing opposing hexagonal spots, and calculating the change in angle between these two lines. Close examination of the moir\'{e} pattern reveals three types of high-symmetry regions highlighted by the circles, labeled 1-3. A magnified view of these regions is shown below Figure~\ref{fig2}(a). The labels AA$'$, AB Nitrogen, and AB Boron will be explained later.
Before proceeding, it is worth explicitly defining some terminology that we will adopt regarding the high-symmetry points that occur in moir\'{e} lattices. Following the terminology of Hermann \cite{hermannPeriodicOverlayersMoire2012}, we will refer to the high-symmetry points highlighted in Figure~\ref{fig2}(a)1 as 'moirons' or nodes. Moirons (nodes) will be distinguished from the other two high-symmetry points highlighted in Figure~\ref{fig2}(a)2-3, which we will refer to as anti-nodes. 

There are two possibilities for stacking order, where the atoms stack directly on top of one another. The first is AA stacking, where the second layer is repeated without translation or rotation relative to the base layer. The second is AA$'$ stacking, where the second layer has a rotation of 180 degrees with respect to the base layer. If there were only one atomic species, these two stacking arrangements would be identical, but h-BN contains a B and N in the unit cell so that these arrangements produce different stacking. These two arrangements are shown in Figure~\ref{fig2}(b) and (c) along with a sketch of different individual and summed intensities. For the sketch, we assume an atom is approximated by a Gaussian intensity distribution with an amplitude equal to z$^2$ ($B = 5^2 = 25$; $N = 7^2 = 49$) \cite{pennycookZcontrastStemMaterials1989,pennycookHighresolutionZcontrastImaging1991}. For AA$'$ stacking, shown in Figure~\ref{fig2}(b) both atomic columns contain one B and one N atom. The two B atoms are represented in the intensity sketch by the blue curves, and the two N atoms by the red curves. Summing these distributions gives the black curve. Here, the column intensities are equivalent. 

We compare this with AA stacking in Figure~\ref{fig2}(c). In this case, one atomic column has two N atoms, while the other atomic column has two B atoms. The intensities of the single atoms are again shown with the blue and red curves, while their sum is shown by the black curve. We see that the two atomic columns have different intensity and that the column with two B atoms has roughly the same intensity as a single N atom.

It is important to stress that Figure~\ref{fig2}(b) and (c) is not intended to be a quantitative simulation or prediction of observed intensities. It is intended to be an illustrative conceptual sketch.

Next, we examine what happens when one layer is twisted by 8 degrees relative to the other for the two cases, AA$'$ and AA stacking. These are shown in Figure~\ref{fig2}(d) and (e). The familiar moir\'{e} pattern is evident and there emerge three distinct stacking arrangements at the high-symmetry points. For AA$'$ stacking, when a rotation is added (shown in \ref{fig2}(d)), we obtain AA$'$ within the moirons and two types of AB stacking in the anti-nodes. For one AB stacking arrangement, B appears on top of B; for the other AB stacking arrangement, N appears on top of N. These two variations of AB stacking are highlighted in the magnified views below the figure. The BB and NN stacks are circled in blue. We will refer to the AB stacked arrangement where B appears on top of B 'AB Boron' and where N appears on top of N 'AB Nitrogen'.

Let us consider how these configurations will appear in our image. We have noted already that a stack of two B atoms will have approximately the same intensity as a single N atom (see Figure~\ref{fig2}(c)). This indicates that for AB Boron, both the stacked B and the unstacked N will have the same intensity. Without sufficient resolution, this region should have approximately uniform intensity.

The situation is different for the AB Nitrogen case. In this case, we have stacked N atoms and unstacked B atoms. We expect to see a large difference in image intensity between stacked N and unstacked B. Moreover, because the stacked N atoms are further apart than the average interatomic distance in this region, we expect to be able to resolve the stacked N atoms even if we do not have sufficient resolution to fully resolve the rest of the structure.

Based on these observations we have good grounds to suggest the labeling AA$'$, AB Nitrogen, and AB Boron for the three high-symmetry points highlighted in our image (Figure~\ref{fig2}(a)).

Let us consider the alternative. If we begin with AA stacking and add a rotation we will maintain AA stacking at some of the nodes. At the other two nodes we will again get a version of AB stacking (either AB or BA) but in this case both versions result in either B on N or N on B with the surrounding unstacked atoms being a mix of B and N. Because these stacking arrangements result in alternating atomic species\textemdash as is the case for AA$'$ stacking\textemdash we use the shorthand AB$'$ stacking to highlight the difference between this case and the AB stacking that we have called AB Boron and AB Nitrogen.

Multislice image simulations were performed to obtian a more quantitative prediction of the relative intensity levels and appearance of the image. Figure~\ref{fig2}(f) and (g) show the AA$'$ and AA stacked case (respectively) with an 8 degree rotation. The nodes of interest are highlighted and magnified views are displayed on the left for comparison with the nodes from the original image. The AB Nitrogen node unambiguously matches the features of the experimental image. Thus, we conclude that our material is AA$'$ stacked with an 8 degree rotation. We can now reliably identify the moirons, and two types of anti-nodes directly from image intensity.

\subsection{Mathematical Representation of Moir\'{e} Patterns from Two Hexagonal Lattices}
\label{sec:moire_math}

Readers already familiar with the mathematics of moir\'{e} lattices may skip sections~\ref{sec:moire_general} and~\ref{sec:moire_self} and proceed directly to section~\ref{sec:moire_hexagonal}. However, the present author found it very helpful to refer to these derivations when thinking about moir\'{e} structures during the analysis. Thus, reproducing them here in condensed form is intended to act as an easy reference to the exact equations within the same document, without the reader needing to consult textbooks or search for other papers.

\subsubsection{Presentation of a General Moir\'{e} Pattern from 2D Lattices}
\label{sec:moire_general}
The general derivation of moir\'{e} patterns from overlapping periodic lattices can be found in reference  \cite{hermannPeriodicOverlayersMoire2012}  and a further expanded discussion can be found in reference  \cite{hermannCrystallographySurfaceStructure2016}. In this section, we will pull from those resources and provide a brief presentation of the mathematics involved. The following will not be a derivation, to be followed step by step, but will be intended to indicate the form that the equations for moir\'{e} patterns take, along with the general idea behind their origin. Once we have presented the general form of the equations, we will make simplifying substitutions for the application to a hexagonal lattice forming a moir\'{e} pattern with an identical but rotated hexagonal lattice. Readers interested in more detail are referred to the previously cited publications. 

These lattices can be described using the Wood notation, $(p_1 \times p_2)R\alpha$ \cite{woodVocabularySurfaceCrystallography1964,hermannCrystallographySurfaceStructure2016}. Here, $\alpha$ refers to the rotation angle between the two lattices, and $p_1$ and $p_2$ represent scaling factors that describe how the overlayer scales relative to the base layer in the two lattice directions. Because the moir\'{e} phenomenon arises due to interference between the two periodic lattices, all of the details regarding the moir\'{e} pattern can be described using relational quantities (i.e.
 scaling and rotation of the overlayer relative to the base layer). This fact makes the mathematical description of the emergent pattern far more manageable than if it were necessary to make reference to the atomic coordinates of each layer. Nevertheless, the general treatment can still be overly cumbersome in situations where obvious simplifications are at hand. Notably, we can see that $p_1$ and $p_2$ will be set to one whenever we have a moir\'{e} pattern that is a result of interference between two lattices of the same periodicity (e.g.
 twisted bilayer graphene and, here, h-BN). It will be helpful to have a concise presentation that leverages these simplifications.

Following the general formalism laid out by Hermann \cite{hermannPeriodicOverlayersMoire2012} we begin by describing a general 2D lattice that has basis vectors $\vec{R_1}$ and $\vec{R_2}$. The reciprocal lattice vectors, $\vec{G_1}$ and $\vec{G_2}$, can be derived from real-space basis vectors using the orthogonality relation

\begin{equation}
\label{eq1}
    \vec{R_i} \cdot \vec{G_j} = 2 \pi \delta_{ij}
\end{equation}

\noindent where $\delta_{ij}=0$ for $i\ne j$ and $1$ for $i=j$. Because we will end up with three lattices\textemdash substrate (or base layer), overlayer, and moir\'{e}\textemdash we will use $s, o$ and $M$ to denote these layers. Our substrate layer can be represented by an infinite Fourier series using the expression

\begin{equation}
    f_s(\vec{r}) = \sum_{j,k}c^s_{j,k} \exp(i[j\vec{G_1} + k\vec{G_2}] \cdot \vec{r}) \hspace{2mm} \text{,}
\end{equation}

\noindent where $j$ and $k$ are integers. In general, the coefficients, $c^s_{j,k}$, will be complex numbers. We can define a similar expression for the overlayer, denoting its lattice vectors as $\vec{R'_i}$ and reciprocal lattice vectors as $\vec{G'_i}$, which are again related through the orthogonality relation from equation~\ref{eq1}. 

\begin{equation}
    f_o(\vec{r}) = \sum_{j',k'}c^o_{j',k'} \exp(i[j'\vec{G'_1} + k'\vec{G'_2}] \cdot \vec{r})
\end{equation}

\begin{figure}
    \centering
    \includegraphics[width=0.75\linewidth]{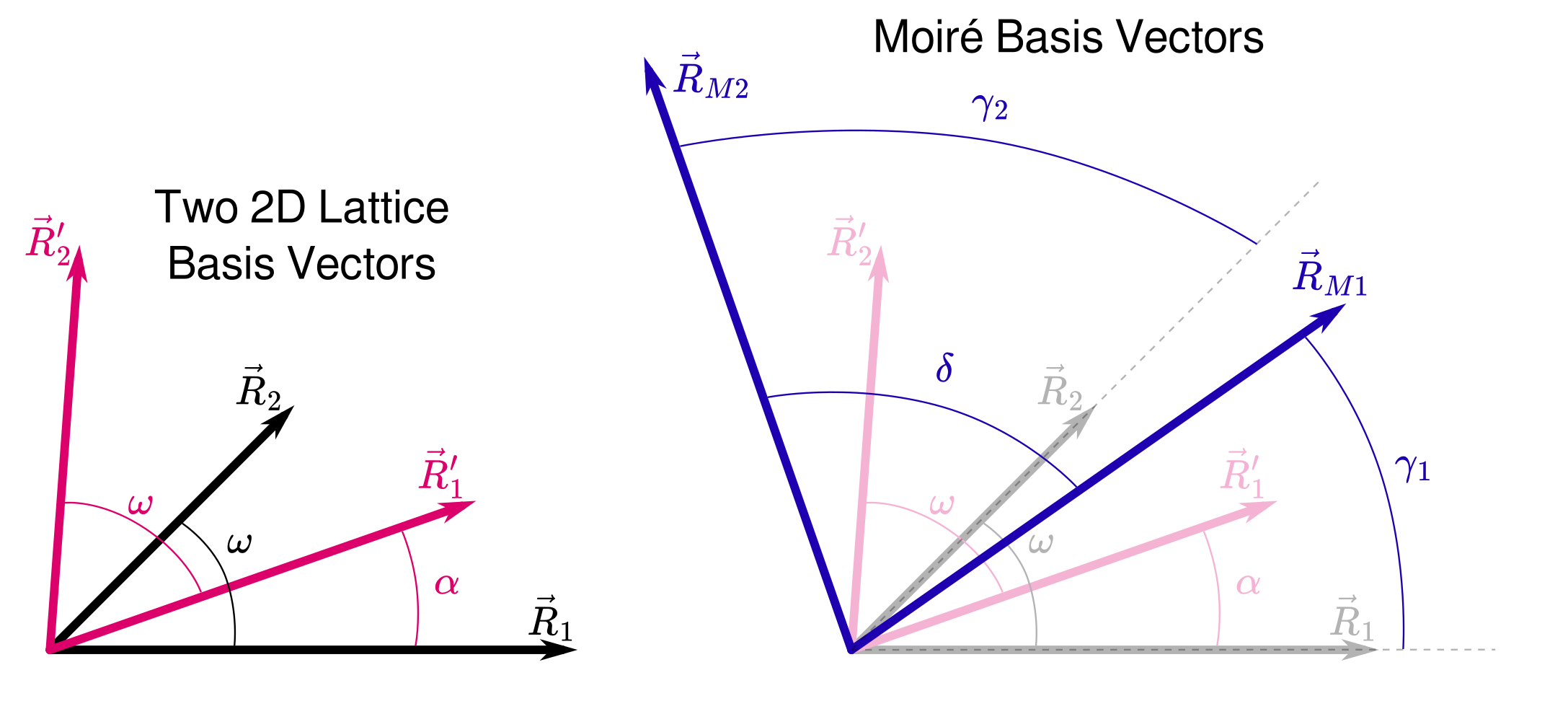}
    \caption{Diagram of the substrate, $\vec{R}_i$, and overlayer, $\vec{R}'_i$, basis vectors and the rotation angle, $\alpha$. The resulting moir\'{e} lattice vectors, $\vec{R}_{Mi}$,  their rotation angles, $\gamma_i$, relative to the basis vectors, and the angle between them, $\delta$.}
    \label{fig3}
\end{figure}

We now have four real space directions\textemdash $\vec{R_1}$ and $\vec{R_2}$ in the substrate and $\vec{R'_1}$ and $\vec{R'_2}$ in the overlayer (refer to figure~\ref{fig3})\textemdash that are related to each other through a linear transformation which may be expressed in matrix notation.

\begin{equation}
\label{eqn4}
    \begin{pmatrix}
        \vec{R'_1}\\
        \vec{R'_2}
    \end{pmatrix}
    = \textbf{M}
    \begin{pmatrix}
        \vec{R_1}\\
        \vec{R_2}
    \end{pmatrix}
\end{equation}

\noindent Here, the bold-face $\textbf{M}$ represents a ($2 \times 2$) transformation matrix that handles both rotation and scaling. We will have a similar expression for the relationship between the reciprocal lattice vectors.

\begin{equation}
\label{eqn5}
    \begin{pmatrix}
        \vec{G'_1}\\
        \vec{G'_2}
    \end{pmatrix}
    = \textbf{K}
    \begin{pmatrix}
        \vec{G_1}\\
        \vec{G_2}
    \end{pmatrix}
\end{equation}

The moir\'{e} pattern arises due to the superpositioning of $f_s$ and $f_o$. By summing over these expressions we expect to find a representation of the moir\'{e}.

\begin{equation}
        f_M(\vec{r}) = f_s(\vec{r}) + f_o(\vec{r}) = \sum_{j,k} c^s_{j,k} \exp(i[j\vec{G_1} + k\vec{G_2}]\cdot \vec{r}) a^M_{j,k}(\vec{r})
\end{equation}

\noindent The factors $a^M_{j,k}$ are also periodic and take the form

\begin{equation}
    a^m_{j,k}(\vec{r}) = 1 + \frac{c^o_{j,k}}{c^s_{j,k}} \exp(i[j(\vec{G'_1} - \vec{G_1}) + k(\vec{G'_2} - \vec{G_2})]\cdot \vec{r})
    \hspace{2mm} \text{.}
\end{equation}
We can see that the difference in the reciprocal lattice vectors, $\vec{G'_i} - \vec{G_i}$, now takes the same role in the exponential as the original reciprocal lattice vectors. These are our moir\'{e} reciprocal lattice vectors and we may define $\vec{G}_{Mi} = \vec{G'}_i - \vec{G}_i$ or

\begin{equation}
    \begin{pmatrix}
        \vec{G}_{M1}\\
        \vec{G}_{M2}
    \end{pmatrix}
     = 
    \begin{pmatrix}
        \vec{G'}_1 - \vec{G}_1\\
        \vec{G'}_2 - \vec{G}_2
    \end{pmatrix}
    \hspace{2mm} \text{.}
\end{equation}

Using equation~\ref{eqn5} we may relate this back to our substrate reciprocal lattice vectors using the expression

\begin{equation}
    \begin{pmatrix}
        \vec{G}_{M1}\\
        \vec{G}_{M2}
    \end{pmatrix}
     = 
     [\textbf{K} - \textbf{1}]\
         \begin{pmatrix}
        \vec{G}_{1}\\
        \vec{G}_{2}
    \end{pmatrix}
\end{equation}

\noindent where $\textbf{1}$ refers to the ($2 \times 2$) identity matrix. Now that we have an expression for our reciprocal moir\'{e} lattice vectors, we may define real space moir\'{e} lattice vectors $\vec{R}_{M1}$ and $\vec{R}_{M2}$ that are derived from the orthogonality relations (equation~\ref{eq1}) applied to the reciprocal moir\'{e} vectors, $\vec{G}_{M1}$ and $\vec{G}_{M2}$. We will then obtain a relation between the substrate lattice vectors and the moir\'{e} lattice vectors that can be expressed as follows.

\begin{equation}
    \begin{pmatrix}
        \vec{R}_{M1}\\
        \vec{R}_{M2}
    \end{pmatrix}
    = \textbf{P}
    \begin{pmatrix}
        \vec{R_1}\\
        \vec{R_2}
    \end{pmatrix}
\end{equation}

We now have a transformation matrix associated with each pair of lattice vectors that can be used to convert from the substrate to the overlayer, from real space to reciprocal space, and from the substrate to the moir\'{e}. There are a number of useful relationships between these various transformation matrices that enable one to calculate $\textbf{K}$ from $\textbf{M}$ and $\textbf{P}$ from $\textbf{K}$.

\begin{equation}
    \textbf{MK}^\top = \textbf{1}
    \hspace{3mm}
    \Rightarrow
    \hspace{3mm}
    \textbf{K} = (\textbf{M}^{-1})^\top = (\textbf{M}^\top)^{-1}
\end{equation}
\begin{equation}
    \textbf{P} [\textbf{K}-\textbf{1}]^\top = \textbf{1}
    \hspace{3mm}
    \Rightarrow
    \hspace{3mm}
    \textbf{P} = ([\textbf{K}-\textbf{1}]^\top)^{-1} = (\textbf{1} - \textbf{M})^{-1}\textbf{M}
\end{equation}
Here, $\textbf{K}^\top$ refers to the transpose; $\textbf{K}^{-1}$ refers to the inverse; $\textbf{1}$ refers to the identity matrix.

Next, we would like to express these transformation matrices in terms of the previously mentioned Wood notation. For a general moire pattern formed from two overlapping lattices we may write $(p_1 \times p_2)R\alpha$, to symbolize the two scaling factors, $p_1$ and $p_2$, and the rotation angle, $\alpha$, that transform the substrate lattice vectors into the overlayer lattice vectors. Specifically,

\begin{equation}
\label{eqn13}
    |\vec{R'}_1| = p_1 |\vec{R}_1|
    \hspace{3mm}
    \Rightarrow
    \hspace{3mm}
    p_1 = \frac{|\vec{R'}_1|}{|\vec{R}_1|}
\end{equation}
and
\begin{equation}
\label{eqn14}
    |\vec{R'}_2| = p_2 |\vec{R}_2|
    \hspace{3mm}
    \Rightarrow
    \hspace{3mm}
    p_2 = \frac{|\vec{R'}_2|}{|\vec{R}_2|} \hspace{2mm} \text{.}
\end{equation}

\noindent In terms of $p_1$, $p_2$, and $\alpha$, we may express the transformation matrix $\textbf{M}$, which allows us to represent the overlayer basis vectors in terms of the substrate basis vectors (equation~\ref{eqn4}), as follows.

\begin{equation}
    \textbf{M} = \frac{1}{\sin{\omega}}
    \begin{pmatrix}
        p_1 \sin(\omega - \alpha) && q p_1\sin{\alpha}\\
        -\frac{1}{q}p_2 \sin{\alpha} && p_2 \sin(\omega + \alpha)
    \end{pmatrix}
    \hspace{2mm} \text{,} \hspace{20mm}
    q = \frac{|\vec{R_1}|}{|\vec{R_2}|}
\end{equation}

\noindent Here, $\omega$ represents the angle between the two basis vectors in the substrate, e.g.
 60 degrees for a hexagonal lattice.

An expression for $\textbf{P}$ may also be derived in terms of the Wood notation parameters as follows,

\begin{equation}
    \textbf{P} = \frac{1}{\Delta \sin{\omega}}
    \begin{pmatrix}
        p_1[\sin(\omega-\alpha)-p_2 \sin{\omega}] && qp_1 \sin{\alpha} \\
        -\frac{1}{q}p_2 \sin{\alpha} && p_2[\sin(\omega+\alpha)-p_1 \sin{\omega}]
    \end{pmatrix}
\end{equation}
where

\begin{equation}
    \Delta = 1 + p_1p_2 - (p_1 + p_2) \cos{\alpha} + (p_1 - p_2) \cot{\omega} \sin{\alpha}
    \hspace{2mm} \text{.}
\end{equation}
We can express the moir\'{e} lattice vector scaling factors, $\kappa_1$ and $\kappa_2$, in terms of the substrate lattice vectors as

\begin{equation}
    \kappa_1 = \frac{|\vec{R}_{M1}|}{|\vec{R}_1|} = \frac{p_1}{\Delta}\sqrt{1+p_2^2 - 2p_2 \cos{\alpha}}
\end{equation}
and

\begin{equation}
    \kappa_2 = \frac{|\vec{R}_{M2}|}{|\vec{R}_2|} = \frac{p_2}{\Delta}\sqrt{1+p_1^2 - 2p_1 \cos{\alpha}}
    \hspace{2mm} \text{.}
\end{equation}
The moir\'{e} rotation angle, $\gamma_1$, represents the angle between the vectors $\vec{R}_{M1}$ and $\vec{R}_1$. The following two expressions relate this angle to the Wood notation parameters.

\begin{align}
    \cos{\gamma_1} &= \frac{\vec{R}_{M1} \cdot \vec{R}_1}{|\vec{R}_{M1}||\vec{R}_1|}\\
     &= \frac{\cos{\alpha}-p_2}{\sqrt{1+p_2^2-2p_2 \cos{\alpha}}}\\
     \tan{\gamma_1} &= \frac{\sin{\alpha}}{\cos{\alpha} - p_2}
 \end{align}
 The moir\'{e} rotation angle, $\gamma_2$, represents the angle between the vectors $\vec{R}_{M2}$ and $\vec{R}_2$. The following two expressions relate this angle to the Wood notation parameters.

 \begin{align}
     \cos{\gamma_2} &= \frac{\vec{R}_{M2} \cdot \vec{R}_2}{|\vec{R}_{M2}||\vec{R}_2|}\\
     &= \frac{\cos{\alpha} - p_1}{\sqrt{1+p_1^2-2p_1 \cos{\alpha}}}\\
     \tan{\gamma_2} &= \frac{\sin{\alpha}}{\cos{\alpha}-p_1}
 \end{align}
 Finally, the angle, $\delta$, between the moir\'{e} lattice vectors $\vec{R}_{M1}$ and $\vec{R}_{M2}$ can be expressed as

 \begin{align}
\cos{\delta} &= \frac{\vec{R}_{M1} \cdot \vec{R}_{M2}}{|\vec{R}_{M1}||\vec{R}_{M2}|} \\
&= \frac{[1+p_1 p_2 - (p_1 + p_2) \cos{\alpha}] \cos{\omega} + (p_2 - p_1) \sin{\alpha} \sin{\omega}}{\sqrt{(1+ p_1^2 - 2p_1 \cos{\alpha})(1 + p_2^2 - 2p_2 \cos{\alpha})}} \\
\tan{\delta} & = \frac{\Delta \tan{\omega}}{1 + p_1 p_2 - (p_1 + p_2) \cos{\alpha} + (p_2 - p_1) \tan{\omega} \sin{\alpha}}
\hspace{2mm} \text{.}
 \end{align}

\subsubsection{Simplifications for a 2D Lattice Forming a Moir\'{e} with Itself}
\label{sec:moire_self}
We now have at hand general expressions for the relationship between key features of the 2D moir\'{e} pattern in terms of the other two layers. Given that we want to examine the moir\'{e} pattern formed by twisting two layers of the same material (h-BN), we know that $|\vec{R}_1| = |\vec{R'}_1|$ and $|\vec{R}_2| = |\vec{R'}_2|$. In words, the magnitude of the basis vectors of both layers is the same because it is the same material (it is only the directions that will be different). This means that $p_1 = p_2 = 1$, from equations \ref{eqn13} and \ref{eqn14}.

Because of the frequency with which moir\'{e} patterns are formed in this way, it is worth rewriting the last few equations with this substitution.

\begin{equation}
    \textbf{M} = \frac{1}{\sin{\omega}}
    \begin{pmatrix}
        \sin(\omega - \alpha) && q \sin{\alpha}\\
        -\frac{1}{q} \sin{\alpha} && \sin(\omega + \alpha)
    \end{pmatrix}
    \hspace{2mm} \text{,} \hspace{20mm}
    q = \frac{|\vec{R_1}|}{|\vec{R_2}|}
\end{equation}
\begin{equation}
    \textbf{P} = \frac{1}{2(1- \cos{\alpha}) \sin{\omega}}
    \begin{pmatrix}
        \sin(\omega-\alpha)-\sin{\omega} && q\sin{\alpha} \\
        -\frac{1}{q} \sin{\alpha} && \sin(\omega+\alpha)- \sin{\omega}
    \end{pmatrix}
\end{equation}
\begin{equation}
    \kappa_1 = \kappa_2 = \kappa = \frac{1}{2|\sin{\frac{\alpha}{2}}|}
\end{equation}
\begin{equation}
\label{rotation}
    \gamma_1 = \gamma_2 = \gamma = 90^{\circ} + \frac{\alpha}{2}
\end{equation}
\begin{equation}
    \cos{\delta} = \cos{\omega}
    \hspace{5mm} \Rightarrow
    \hspace{5mm}
    \delta = \omega
\end{equation}
For a moir\'{e} pattern obtained by self interference, the moir\'{e} lattice we will obtain has a periodicity that scales inversely with the sine of half the rotation angle ($\kappa$). As the rotation angle goes to zero, the periodicity diverges.
 
Any crystallographic direction in the substrate lattice has a similar direction in the moir\'{e} pattern that may be found by rotating by $90^{\circ}$ plus half the rotation angle ($\gamma$).

The angle between the moir\'{e} basis vectors will be the same as the angle between the original basis vectors ($\delta$).

\subsubsection{Simplifications for a Hexagonal 2D Lattice Forming a Moir\'{e} with Itself}
\label{sec:moire_hexagonal}

For a hexagonal lattice, specifically, we have the further simplification that $q = 1$ and $\omega = 60^{\circ}$, or $\pi /3$ in radians. A few more substitutions become possible and we can express all relations in terms of only the rotation angle, $\alpha$.

\begin{equation}
    \textbf{M} = \frac{2}{\sqrt{3}}
    \begin{pmatrix}
        \sin(\pi /3 - \alpha) && \sin{\alpha}\\
        -\sin{\alpha} && \sin(\pi /3 + \alpha)
    \end{pmatrix}
\end{equation}
\begin{equation}
    \textbf{P} = \frac{1}{\sqrt{3}(1- \cos{\alpha}) }
    \begin{pmatrix}
        \sin(\pi / 3-\alpha)- \frac{\sqrt{3}}{2} && \sin{\alpha} \\
        -\sin{\alpha} && \sin(\pi / 3+\alpha)- \frac{\sqrt{3}}{2}
    \end{pmatrix}
\end{equation}
\begin{equation}
    \kappa_1 = \kappa_2 = \kappa = \frac{1}{2|\sin{\frac{\alpha}{2}}|}
\end{equation}
\begin{equation}
    \gamma_1 = \gamma_2 = \gamma = 90^{\circ} + \frac{\alpha}{2}
\end{equation}
\begin{equation}
    \delta = \pi / 3
\end{equation}
 
\subsection{E-Beam Milling in Twisted Bilayer h-BN}
\label{sec:twisted_milling}
So far, we have closely examined the HAADF image features for twisted bilayer h-BN and provided a condensed mathematical treatment of moir\'{e} patterns for bilayer hexagonal lattices. This mathematical analysis allows us to translate information about the moir\'{e} pattern into information about the alignment of the two lattices. With these tools in hand, we can begin to examine the results of e-beam milling in twisted bilayer h-BN.

\subsubsection{Labeling Directions in the Moir\'{e} Pattern}
\label{sec:moire_directions}
\begin{figure}
\centering
  \includegraphics[width=\linewidth]{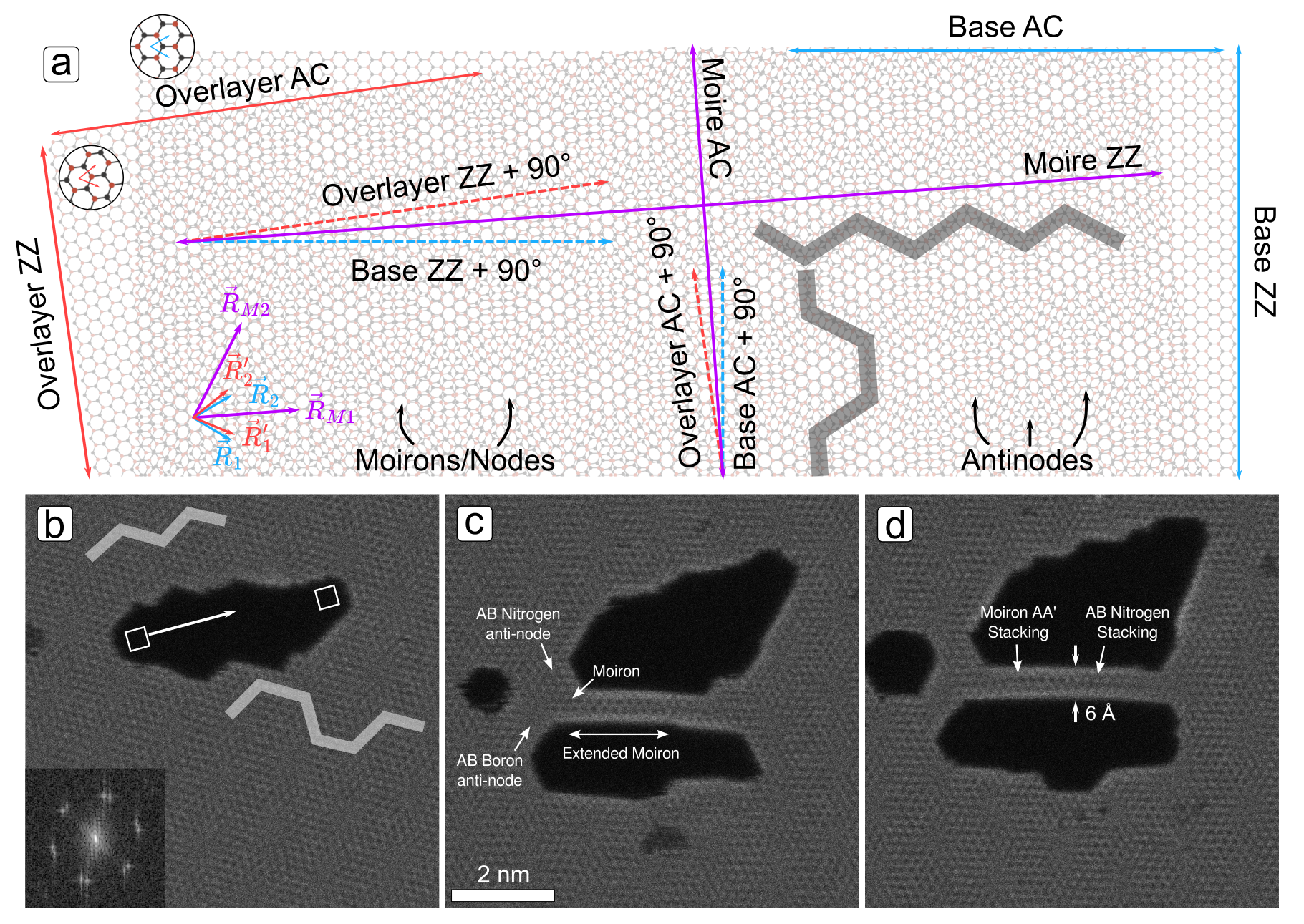}
  \caption{Summary of milling results in twisted bilayer h-BN. (a) Diagram illustrating the AC and ZZ directions in the two layers and the moir\'{e}. (b) HAADF image of the result of the first cut in the ZZ direction. (c) Additional cut in the AC direction. (d) Further milling produces an atomically sharp h-BN nanoribbon that is 6 \AA{} wide. }
  \label{fig4}
\end{figure}

Figure~\ref{fig4}(a) shows an example of a moir\'{e} pattern formed from two hexagonal lattices, with the overlayer rotated by $8^{\circ}$. The moirons tend to be, visually, what we might call the primary feature of the moir\'{e} pattern. For this reason, it is tempting to liken them to the atoms in the original lattices, but this is a mistake. The moirons form a close packed hexagonal pattern which should be likened to the negative space between the atoms in the original lattice. Around each moiron we have a hexagon of anti-nodes. As we have seen already (Figure~\ref{fig2}(a), (d), and (f)) these anti-nodes are of two different, alternating types. It is not the moirons, but the anti-nodes that form the analog of the atoms in the moir\'{e} pattern.

This becomes clearer when we consider assigning analogous directions such as ZZ and AC to the moir\'{e} pattern. The mathematical analysis suggested that for any direction the corresponding direction in the moir\'{e} pattern will be found by rotating $90^{\circ} + \alpha/2$. Figure~\ref{fig4}(a) shows a schematic diagram illustrating the zig-zag (ZZ) and arm-chair (AC) directions. The overlayer ZZ (AC) direction is rotated by 8 degrees relative to the base ZZ (AC) direction. However, the moir\'{e} ZZ (AC) direction is rotated by roughly $90^{\circ}$ relative to the other two. We can think of this as taking the average of a $90^{\circ}$ rotation with respect to the base layer and the over layer, or `splitting the difference' ($\alpha/2$) between the two. Thus, the AC and ZZ directions in the moir\'{e} pattern are AC and ZZ relative to the anti-nodes, not the moirons.

\subsubsection{Milling Along ZZ and AC Moir\'{e} Directions}
\label{sec:twisted_milling_results}
In the milling experiment shown in Figure~\ref{fig4}(b) a small subscan box was defined and the e-beam was confined to scan within this box. This enabled the dose to be localized to a small area of the sample, which was then translated within the larger imaging area (simultaneous imaging at both scales is not possible). As atoms were ejected from the sample, the subscan box was translated iteratively to produce a cut in the h-BN. The initial milling direction chosen was the moir\'{e} ZZ direction. The result of this procedure is shown in Figure~\ref{fig4}(b). The edges of the cut are very rough and it appears that the AC direction is preferential---fairly uniform, flat regions have appeared that are aligned with the AC direction and none in the ZZ direction.

The scan direction was then rotated to align with the AC direction and further milling was performed in an attempt to see if atomically precise shaping could be achieved. The results of this attempt are shown in Figure~\ref{fig4}(c) and (d). A 6 \AA{} wide nanoribbon was able to be formed with atomically smooth edges.

A few features observed in this ribbon are worth highlighting. In Figure~\ref{fig4}(c), on the left side of the ribbon, we see the AB Nitrogen anti-node on top and the AB boron anti-node on the bottom. The moiron begins as expected, at the left edge of the nanoribbon, but persists along the nanoribbon core much further than would be expected due to the moir\'{e}. Likely the edge reconstruction during the milling process allowed the $8^{\circ}$ rotation angle to be undone, resulting in an AA$'$ stacked nanoribbon core.

After further exposure to the e-beam (Figure~\ref{fig4}(d)), we observe that the left half of the nanoribbon retains the moiron AA$'$ stacking while the right side of the nanoribbon appears to adopt the AB nitrogen anti-node stacking. The bright edges of the nanoribbon suggest that the edges are either terminated with nitrogen on both sides (N-N stacking) or that more than two atoms are involved in the edge reconstruction (e.g.
 N-B-B, N-N-B etc.). Both of these possibilities would result in a higher intensity than what is observed in the bulk of the nanoribbon.

\section{AA$'$ Stacked h-BN}
\label{sec:aa_prime}
\subsection{Testing the Role of Twist Angle}
\label{sec:testing_twist}
The ability to create an atomically sharp nanoribbon edge in twisted bilayer h-BN is perhaps somewhat surprising given the stochastic nature of the atomic ejection process. This success raises a fundamental question: is the twist angle itself necessary for achieving atomic precision, or is it the crystallographic orientation with respect to the moir\'{e} lattice that matters? To address this question, we examine milling in an AA$'$ stacked h-BN bilayer without a twist.

At first glance, discussing moir\'{e} directions in an untwisted bilayer may seem contradictory---without rotation, there is no moir\'{e} pattern. However, the mathematical framework developed in Section~\ref{sec:moire_math} reveals that the untwisted case represents the limit as the rotation angle approaches zero ($\alpha \rightarrow 0$). In this limit, the moir\'{e} periodicity diverges to infinity, but the relationship between crystallographic directions remains well-defined. According to Equation~\ref{rotation}, any crystallographic direction in the moir\'{e} pattern is rotated by $90^{\circ} + \alpha/2$ relative to the same direction in the substrate. For $\alpha = 0$, this becomes exactly $90^{\circ}$. Therefore, the moir\'{e} armchair (AC) direction---which produced atomically sharp edges in the twisted case---corresponds to the zigzag (ZZ) direction in the substrate for an untwisted bilayer.

This mathematical insight provides a clear experimental test: if the moir\'{e} crystallographic direction is the critical factor rather than the twist angle itself, then milling along the substrate ZZ direction in untwisted AA$'$ h-BN should produce results comparable to milling along the moir\'{e} AC direction in twisted bilayer h-BN.

\subsection{Initial Milling Attempts and an Unexpected Confound}
\label{sec:initial_attempts}

We began by attempting to mill along the substrate AC direction, initially thinking this would be analogous to the successful moir\'{e} AC direction from the twisted bilayer experiments. However, as the mathematical analysis makes clear, this was incorrect---the substrate AC direction corresponds to the moir\'{e} ZZ direction, which had produced rough edges in the twisted case. For this initial attempt, we defined a long subscan box representing the entire target milling region and exposed it continuously. This approach, which we term ``parallel milling,'' differs from the ``sequential milling'' approach used in the twisted bilayer experiments, where a small subscan box was translated incrementally. At the time, we did not consider that this methodological difference might influence the milling quality.

Figure~\ref{fig5}(a) shows the result of this parallel milling attempt along the substrate AC direction. The cut exhibits rough edges with only occasional regions of atomic smoothness---consistent with milling along an unfavorable crystallographic direction (moir\'{e} ZZ equivalent).

We then performed a second milling experiment after rotating the sample to align with the substrate ZZ direction---the predicted favorable orientation corresponding to moir\'{e} AC. The results, shown in Figure~\ref{fig5}(b), show improvement with more regions exhibiting atomically sharp edges. However, the overall quality remains inferior to the twisted bilayer results shown in Figure~\ref{fig4}. The surrounding material shows significant beam damage, which may contribute to the reduced edge quality. Nevertheless, the improvement when switching from substrate AC to substrate ZZ supports the hypothesis that moir\'{e} crystallographic direction matters.

However, this explanation is not fully satisfactory: if the moir\'{e} direction is the critical factor, why were the untwisted results noticeably worse than the twisted results when both were milled along the favorable moir\'{e} AC direction?

\subsection{Isolating the Effect of Milling Strategy}
\label{sec:isolating_strategy}

Reviewing the experimental procedures, we realized that the twisted and untwisted experiments differed not only in twist angle but also in milling strategy. The twisted bilayer had been milled sequentially (small moving subscan box), while the untwisted attempts used parallel milling (large stationary subscan box). To isolate this variable, we performed a third experiment: sequential milling along the substrate ZZ direction (moir\'{e} AC equivalent) in the untwisted AA$'$ bilayer.

The results, shown in Figure~\ref{fig5}(c), are dramatically improved. The cut now exhibits atomically sharp or near-atomically-sharp edges comparable in quality to the twisted bilayer results. The large deviation on the bottom edge can be attributed to milling through a defect patch visible in the previous image. Excluding this defect region, the edge quality rivals that achieved in the twisted bilayer system.

This result resolves the puzzle and leads to two important conclusions. First, the twist angle itself is not necessary for achieving atomic precision---what matters is milling along the appropriate crystallographic direction relative to the moir\'{e} lattice (or its mathematical equivalent in the untwisted limit). Second, and unexpectedly, the milling strategy---sequential versus parallel---plays a critical role in determining edge quality. This second discovery was unanticipated and warrants detailed investigation, which we undertake in Section~\ref{sec:milling_model}.

\subsection{Edge Structure and Reconstruction}
\label{sec:edge_structure}

\begin{figure}
\centering
  \includegraphics[width=\linewidth]{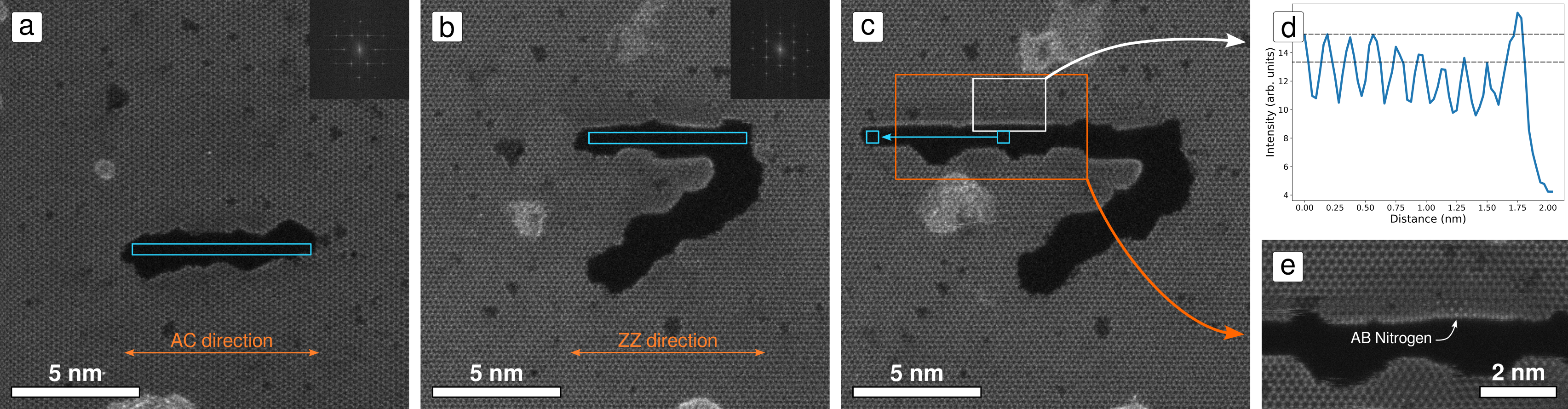}
  \caption{Summary of milling results in AA' stacked bilayer h-BN. (a) Initial cut in the AC direction. Blue box indicates the parallel milling location. (b) Second cut in the ZZ direction. (c) Third cut using a sequential milling strategy. Blue boxes indicate the start and end of the cut. (d) Intensity profile near the edge of the cut, showing a decrease in intensity approaching the edge and then an increase in intensity at the edge. (e) Magnified view of the edge. Beam damage is already evident. }
  \label{fig5}
\end{figure}

To better understand the atomic-scale structure of the milled edges, we obtained higher magnification images of the sequential milling result, shown in Figure~\ref{fig5}(e). Although additional beam damage is already evident in this image---indicating that the original cut was of even higher quality---several important features can be observed.

Figure~\ref{fig5}(d) shows an intensity profile taken from the white boxed region in Figure~\ref{fig5}(c). Horizontal dotted lines serve as guides for the eye. In the bulk region, away from the edge, the intensity corresponds to AA$'$ stacking (alternating N-B and B-N columns). As we approach the edge, the intensity decreases, suggesting a gradual transition from AA$'$ stacking toward AB stacking where atoms are no longer directly aligned. At the edge itself, the intensity increases above the bulk level. This enhanced edge intensity suggests either nitrogen-terminated edges (N-N stacking) or edge reconstructions involving more than two atoms (e.g.
, N-B-B or N-N-B configurations). Both scenarios would produce higher intensity than the AA$'$ bulk stacking.

Interestingly, this intensity pattern is not symmetric. The lower edge of the cut, most clearly visible at the lower left corner of Figure~\ref{fig5}(e), maintains AA$'$ stacking right up to the edge with minimal disturbance to the surrounding material. This asymmetry indicates that atomically precise cuts with minimal collateral damage are achievable, and that specific details---crystallographic position of the edge, local defects, or even scan direction---govern the degree of edge reconstruction. The milling box was scanned from top to bottom, and this directional asymmetry may influence the result. For example, atoms ejected from the bottom edge might preferentially redeposit on the top edge, creating an excess of mobile atoms that facilitate reconstruction.

Another notable feature is the AB nitrogen stacking evident near the top edge in Figure~\ref{fig5}(e). This feature was absent in the earlier image (Figure~\ref{fig5}(c)), indicating that continued beam exposure induced further layer realignment. The transition from AA$'$ to AB nitrogen stacking over such a short distance implies significant in-plane strain, with one layer under compression and the other under tension. The edge deviation from linearity at this location is consistent with this strain distribution and corroborates the gradual intensity decrease observed in Figure~\ref{fig5}(d).

\section{Milling Model: Sequential vs. Parallel}
\label{sec:milling_model}
\subsection{Implementation of Milling Model}
\label{sec:model_implementation}
The experiments in Section~\ref{sec:aa_prime} revealed an unexpected result: the milling strategy---sequential versus parallel---dramatically affects edge quality even when the crystallographic orientation and total milled area are identical. This discovery raises a fundamental question: why should these two approaches, which ostensibly deliver the same total electron dose to the same region, produce such different outcomes?

At first glance, this difference is puzzling. Both methods aim to eject all atoms from the target region, and one might expect that the final result would depend only on the total integrated dose. However, the milling process is goal-based and feedback-driven rather than operating with a predefined dose. The operator continues milling until all atoms in the target region are ejected, and the stochastic details of the ejection process---particularly the exposure of surrounding material to beam tails---can differ significantly between the two approaches.

\begin{figure}
\centering
  \includegraphics[scale=0.6]{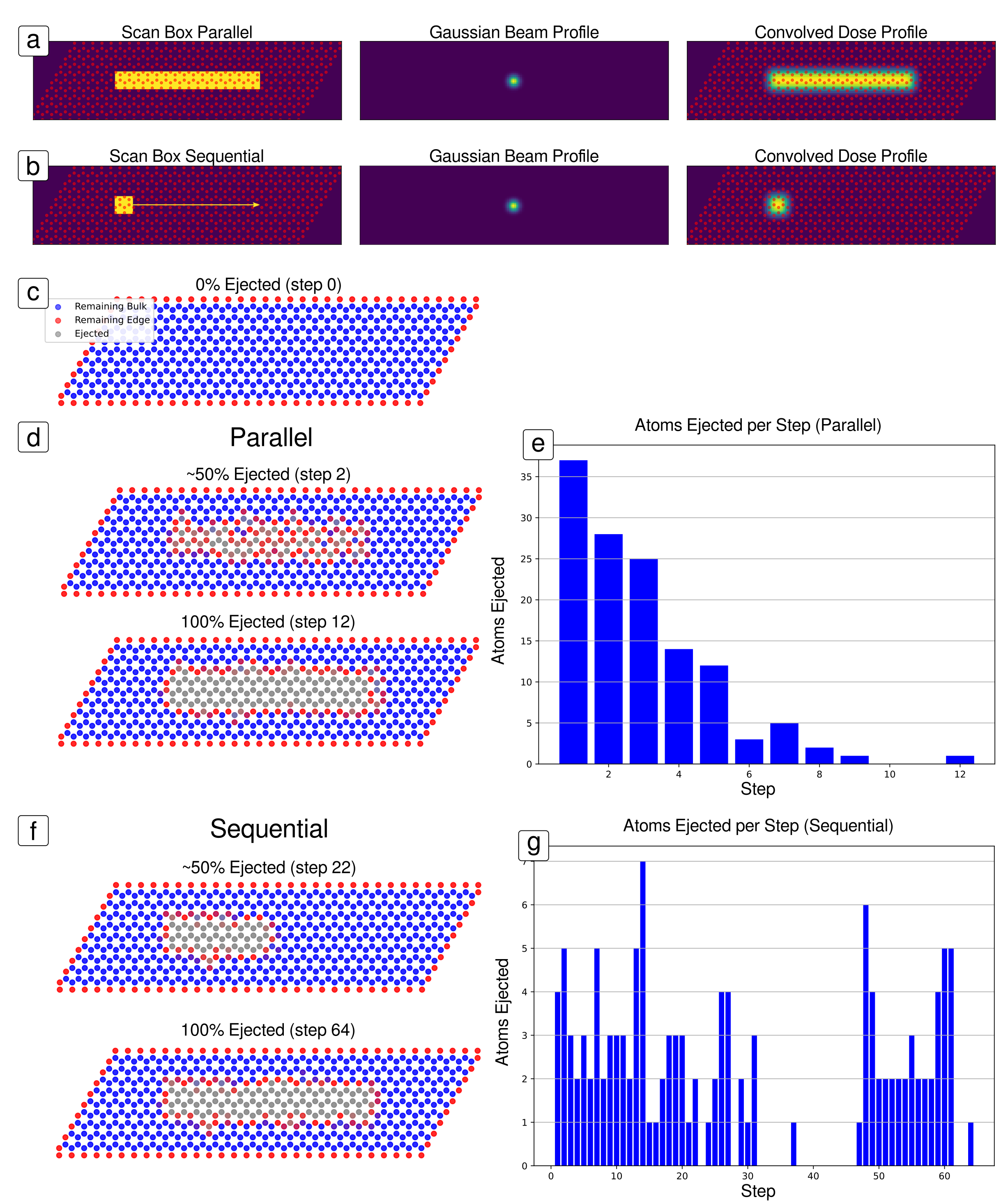}
  \caption{Summary of e-beam milling model. (a) (Left) Schematic of AA' stacked h-BN with target milling region (scan box) highlighted in yellow. (Middle) Gaussian beam profile. (Right) Convolution of the target milling region with the gaussian profile. (b) Same depiction as in (a) but for a square target milling region that will be sequentially advanced during the milling to cover the same area covered in (a). (c) Initial state of the lattice with bulk atoms marked in blue and edge (undercoordinated) atoms marked in red. Ejected atoms (which are shown in (d) and (f)) are marked in gray. (d) Comparison of the lattice state where 50\% of the target atoms have been ejected and 100\% of the target atoms have been ejected, in the parallel milling case shown in (a). (e) Bar chart illustrating the number of atoms ejected at each step, for the parallel milling case. (f) 50\% and 100\% comparison for the sequential milling case. (g) Bar chart illustrating the number of atoms ejected at each step, for the sequential milling case.}
  \label{fig6}
\end{figure}

In this section, we aim to make the previous comment explicit by using a quantitative model of the two milling processes. This model is not aimed at matching the experimental parameters but highlighting the difference between the two milling approaches given some basic assumptions. We model the two approaches as laid out in figure~\ref{fig6}(a) and (b). A bilayer (AA$'$ stacked) h-BN lattice is defined, represented by the red dots. A target area is selected for milling. For the parallel approach, shown in figure~\ref{fig6}(a), the target area is the same as the total scan area and is shown by the yellow box on the left. For the sequential approach, shown in figure~\ref{fig6}(b), the yellow box on the left indicates the scan area for the first step of the milling process. This box will be moved, as the milling progresses, to eventually cover the same area shown in (a). The e-beam is modeled as a 2D Gaussian and is the same for both approaches. Instead of working through the details of a scanning process, we model each scan within the scan box as a single application of the dose by convolving the Gaussian profile with the scan box. The dosed region, after a single `scan', is shown as a heat map on the right for both approaches.

Next, we distinguish between four types of atoms. We will have B and N with different ejection probabilities, but we will also have bulk atoms and edge atoms with different ejection probabilities. In the real world, we would have coordination numbers of 3, 2, and 1 as well as interaction between the layers and dynamic restructuring. For the sake of simplicity we ignore these details and simply assign four ejection probabilities to our four types of atoms ($B_{bulk} = 5\times10^{-21}$, $B_{edge} = 7\times10^{-21}$, $N_{bulk} = 3\times10^{-21}$, $N_{edge} = 5\times10^{-21}$ barns). This scheme is intended to capture the idea that edge atoms eject more quickly than bulk atoms and boron ejects more quickly than nitrogen. We additionally multiply all of these by a constant (in what is presented here we used $5\times10^{13}$). This simply allows us to have a single parameter that can be used to speed up or slow down the simulation so that it completes in a timely manner but does not complete in a single step.

The simulation run for the parallel milling approach consists of applying the convolved dose profile, using a random number generator to decide whether each atom was ejected, flagging its status (ejected or not), and reapplying the dose profile iteratively until all atoms within the milling region have been ejected. At each step, we reassess which atoms are bulk atoms and which are part of the edge (undercoordinated) and adjust their ejection probability accordingly. Notice also: because the dose profile is a convolution of the scan box and the Gaussian beam, regions external to the scan box are exposed to the beam tails. These atoms are also evaluated for ejection with the same process.

For the sequential milling approach, a similar procedure was used except that when all atoms have been ejected from the milling region, the scan box advances by half the width of the scan box (rightward on the page) until it has eventually covered the same region that is covered by the parallel milling scheme. The choice of half the width of the scan box was selected because this roughly approximates what was done experimentally.

Figure~\ref{fig6}(c) shows the initial state of the simulation. Bulk atoms are shown in blue, edge atoms are shown in red, and ejected atoms---none present at this stage---are shown in gray. Although it is not indicated in the diagram, there are two (AA$'$ stacked) h-BN layers here. Figure~\ref{fig6}(d) and (e) show the results of the parallel milling process. In figure~\ref{fig6}(d) we show the state of the layers at $\sim 50\%$ and $100\%$ ejected. Approximately half of the atoms were ejected in the first two dose applications and all atoms were ejected in twelve steps. A histogram of the number of atoms ejected at each step is shown in figure~\ref{fig6}(e). Most atoms are ejected at the beginning. As fewer atoms are present within the dosed area, the number of atoms ejected at each step diminishes.

Figure~\ref{fig6}(f) and (g) show the results of the sequential milling process. Figure~\ref{fig6}(f) shows the state at $\sim 50\%$ and $100\%$ ejected and Figure~\ref{fig6}(g) shows the histogram of the number of atoms ejected at each step. The sequential milling required 64 steps to complete, compared to just 12 for the parallel milling. However, one must be careful with the interpretation of these numbers. They are not equivalent time steps, they are dose application steps. Based on how we have set up the model, the length of time required for scanning the small square used in the sequential milling process would be 1/8 of the time required to scan the larger parallel milling box (it is 1/8 the size of the larger box). So, as far as time is concerned, the parallel milling took the equivalent of 96 sequential milling steps ($12 \times 8 =96$), significantly \textit{longer} than the sequential milling. The sequential milling only ejects about three atoms per step---although this varies stochastically---but the milling rate does not drop off as more atoms are ejected since the scanning box translates to new areas.

\subsection{Examination of the Dose Difference Between Sequential and Parallel Milling}
\label{sec:dose_difference}
\begin{figure}
\centering
  \includegraphics[width=\linewidth]{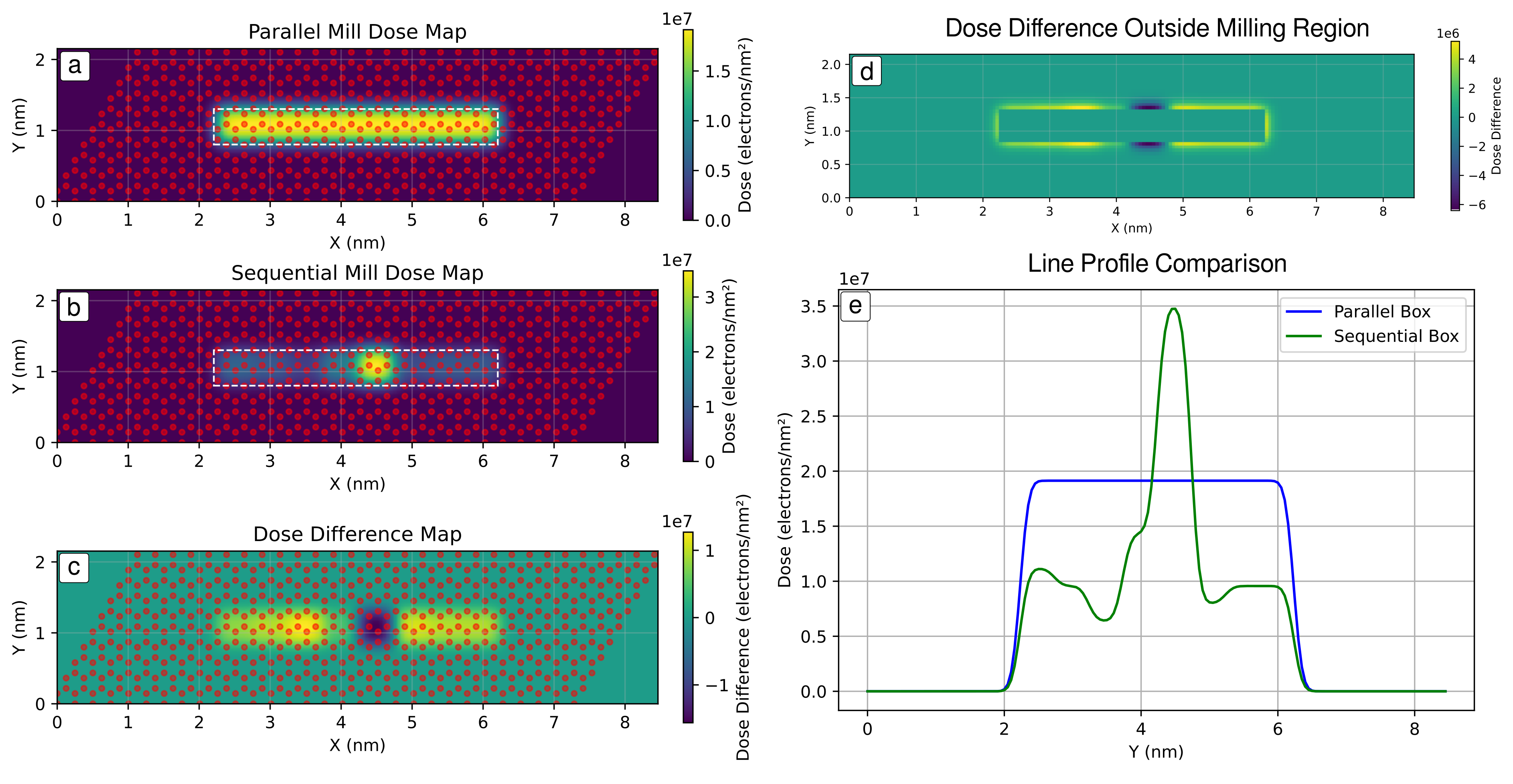}
  \caption{Summary of the dose differences between the parallel and sequential milling strategies. (a) and (b) shown heat maps of the applied dose for the parallel milling and sequential strategy, respectively. (c) Heat map of the difference (Parallel - Sequential) between (a) and (b). (d) Heat map of (c) with the target location set to zero. This captures the difference in dose outside the intended milling region. (e) Line profile comparison of the applied doses from (a) and (b). Line profiles were acquired horizontally across the midpoint of the milling region.}
  \label{fig7}
\end{figure}

Next, we compare the doses required to complete the milling. Figure~\ref{fig7} provides a summary overview of the applied doses. Figure~\ref{fig7}(a) and (b) show heat maps of the dose distribution for the parallel milling and sequential milling approach, respectively. The white, dotted box shows the target milling location, for reference, and the lattice cites are represented by red dots overlaid on top of the heat map. For the sequential milling case, shown in Figure~\ref{fig7}(b), we observe a pronounced variation in dose distribution across the width (horizontally) of the milling region. In particular, we note a relatively high dose at around x = 4.5 nm which correlates with the low ejection rate in Figure~\ref{fig6}(g) between steps 30 and 50. For this milling position, it (randomly) required many steps to eject all the atoms, resulting in a much higher applied dose.

Figure~\ref{fig7}(c) shows a heat map of the difference between the two (parallel minus sequential). In most positions, the sequential approach required less dose. The reader will recall that the purpose of these simulations is to provide grounds for understanding why an empirical difference was observed between the parallel and sequential approaches, particularly with regard to the edge roughness. Since the ejected atoms are the same in both cases, we are really interested in the dose applied outside of our milling region. It is this dose that may provide insight into why the sequential milling approach performed better than the parallel milling approach in the experiments. Figure~\ref{fig7}(d) shows the dose difference heat map with the target milling region set to zero. In other words, this dose map shows only the dose applied outside the intended milling region. We see that, for the majority of the surrounding area, the sequential milling approach resulted in a lower dose. Figure~\ref{fig7}(e) shows a line profile comparison of the heat maps from (a) and (b) taken horizontally across the middle of the milling region.

It should be noted that, because this is a stochastic process, the sequential milling approach is not guaranteed to complete with a lower dose. As is illustrated by the prominent high-dose location, which exceeds the parallel milling dose by almost double, we can have instances where the sequential milling requires more dose. Likewise, the parallel milling strategy could stochastically be completed with a lower dose than what is shown here. However, the key difference is that the parallel milling strategy will result in the maximum dose applied to the whole region, whereas the sequential milling strategy limits the maximum dose to only that region in which the maximum dose was required.

We have deliberately not performed further analysis of the results of this milling model. It is an oversimplification of the real sample dynamics that take place during milling and, thus, should not be assumed to be realistic in the atomic details. We have already seen that the sample reconstructs during milling, involving overlayer rotations, edge reconstructions, and lattice strain. In addition, our ejection probabilites are somewhat arbitrarily chosen and our stochastic approach---deciding whether an atom is ejected or not, based on random number generation---completely ignores the physical energetics and mechanics that govern these processes. The main insight we want to extract from this exercise is simply that, in a milling process like this, with everything else being equal, the sequential milling approach will minimize unnecessary dose to the surrounding specimen. This feature contributes to the sharper edges observed in the sequential milling shown in Figure~\ref{fig5}(c).

\section{Conclusion}
\label{sec:conclusion}
This work establishes a framework for achieving atomic precision in electron beam sculpting of two-dimensional materials by identifying the critical factors that enable controlled nanofabrication. The investigation was motivated by the failure of monolayer h-BN milling to produce atomically sharp edges, leading to the hypothesis that bilayer systems might provide the structural constraints necessary for atomic-scale control.

To interpret milling results in twisted bilayer systems, we leverage systematic analysis of HAADF image contrast combined with multislice simulations to identify different stacking configurations from their distinct intensity signatures at high-symmetry points within the moir\'{e} pattern. This enables reliable identification of AA$'$, AB Nitrogen, and AB Boron stacking arrangements directly from experimental images. We further apply the mathematical framework for moir\'{e} lattices, which establishes that crystallographic directions in the emergent pattern are rotated by $90^{\circ} + \alpha/2$ relative to the substrate, where $\alpha$ is the twist angle. This relationship holds even in the untwisted limit ($\alpha \rightarrow 0$), where the moir\'{e} periodicity diverges to infinity but directional relationships remain well-defined.

With these analytical tools in hand, systematic milling experiments reveal two key findings. First, milling along the moir\'{e} armchair direction produces atomically sharp edges in both twisted and untwisted bilayers, achieving nanoribbons as narrow as 6 \AA{}. Critically, this demonstrates that the twist angle itself is not necessary for atomic precision---rather, it is the crystallographic orientation relative to the moir\'{e} lattice (or its mathematical equivalent) that determines edge quality.

Second, and unexpectedly, the milling strategy plays a decisive role in achieving atomic precision. Sequential milling, where a small electron beam subscan area is translated during the process, produces significantly superior results compared to parallel milling of the entire target region. Stochastic modeling reveals that sequential milling minimizes unwanted exposure to surrounding material through beam tail effects, reducing collateral damage and enabling the formation of atomically smooth edges.

These results advance the state of the art in top-down electron beam nanofabrication and provide insights applicable beyond h-BN to other van der Waals bilayer systems. By demonstrating that atomic precision can be achieved through proper understanding of crystallographic orientation and milling strategy---rather than requiring specific twist angles or exotic material properties---this work opens pathways toward reliable, reproducible atomic-scale manufacturing using focused electron beams.


\section{Experimental Section}
\label{sec:experimental}

Imaging was performed using a Nion UltraSTEM 200 operated at an accelerating voltage of 100 kV, a nominal beam current of 18 pA, and a nominal convergence angle of 32 mrad. Chemical vapor deposition (CVD) grown h-BN was transferred onto Protochips heating chips. Prior to imaging, the sample was baked in vacuum at 160 $^{\circ}$C overnight to remove surface contamination. Although the sample was mounted on a heating chip, this functionality was not used during these experiments.

\medskip
\textbf{Acknowledgements} \par 
This work was supported by the U.S. Department of Energy, Office of Science, Basic Energy Sciences, Materials Sciences and Engineering Division and was performed at the Oak Ridge National Laboratory\'s Center for Nanophase Materials Sciences (CNMS), a U.S. Department of Energy, Office of Science User Facility. Parts of this work was supported by the U.S. Department of Energy, Office of Science, National Quantum Information Science Research Centers, Quantum Science Center (M. B., R.G.M.)

\medskip
\bibliographystyle{unsrt}
\bibliography{bibliography}

\clearpage
\section*{Supplementary Information}
\addcontentsline{toc}{section}{Supplementary Information}

\setcounter{section}{0}
\renewcommand{\thesection}{S\arabic{section}}
\renewcommand{\theHsection}{SI.\arabic{section}}
\setcounter{subsection}{0}
\renewcommand{\thesubsection}{S\arabic{section}.\arabic{subsection}}
\renewcommand{\theHsubsection}{SI.\arabic{section}.\arabic{subsection}}
\setcounter{subsubsection}{0}
\renewcommand{\thesubsubsection}{S\arabic{section}.\arabic{subsection}.\arabic{subsubsection}}
\renewcommand{\theHsubsubsection}{SI.\arabic{section}.\arabic{subsection}.\arabic{subsubsection}}

\setcounter{figure}{0}
\renewcommand{\thefigure}{S\arabic{figure}}
\renewcommand{\theHfigure}{SI.\arabic{figure}}

\setcounter{table}{0}
\renewcommand{\thetable}{S\arabic{table}}
\renewcommand{\theHtable}{SI.\arabic{table}}

\setcounter{equation}{0}
\renewcommand{\theequation}{S\arabic{equation}}
\renewcommand{\theHequation}{SI.\arabic{equation}}


\section{Multislice STEM Simulations of h-BN Bilayers}

\subsection{Motivation and Background}

High-angle annular dark-field (HAADF) imaging in scanning transmission electron microscopy (STEM) produces intensity that is approximately proportional to the square of the atomic number (Z$^2$). For bilayer h-BN with AA$'$ stacking, there are two types of atomic columns: B-N columns (boron atom on bottom layer, nitrogen atom on top layer) and N-B columns (nitrogen atom on bottom layer, boron atom on top layer). A natural question arises: do these two column types produce different intensities in HAADF images? Under a simple Z$^2$ approximation, one might expect identical intensities since both columns contain one boron (Z=5) and one nitrogen (Z=7) atom. However, the order of atoms along the beam direction could potentially affect the intensity through dynamical scattering effects.

Understanding this intensity behavior is critical for interpreting experimental images of twisted bilayer h-BN, where different stacking configurations (AA$'$, AB Nitrogen, AB Boron) produce distinct high-symmetry regions within the moir\'{e} pattern. The ability to identify these stacking configurations from image intensity is essential for understanding the milling results presented in the main text.

\subsection{Simulation Methodology}

Multislice STEM simulations were performed using the py\_multislice package \cite{pymultislice}. The multislice algorithm divides the specimen into thin slices perpendicular to the beam direction and propagates the electron wave function through each slice, accounting for both phase shifts (due to the specimen potential) and diffraction.

\subsubsection{Simulation Parameters}

The microscope parameters used were: accelerating voltage of 100 kV, convergence semi-angle of 30 mrad, HAADF detector range of 80--150 mrad, and defocus varied from $-40$ to $+40$ \AA{} in 11 steps. The computational parameters included a grid size of 512$\times$512 pixels, 40 frozen phonon configurations (unless otherwise noted), and slice positions at 1.8 \AA{} (below first layer) and 5.1 \AA{} (below second layer).

The frozen phonon method accounts for thermal diffuse scattering by running multiple simulations with atoms randomly displaced according to their thermal vibration amplitudes, then averaging the results.

\subsubsection{Structure Files}

Crystal structures were generated using VESTA and exported in .p1 format for compatibility with py\_multislice. The AA$'$ stacked bilayer has a c-axis lattice parameter of 6.66 \AA{}, corresponding to an interlayer spacing of approximately 3.33 \AA{}. Monolayer h-BN was simulated with c = 3.33 \AA{} for comparison.

\subsection{Results: Defocus Series}

\begin{figure}[h]
\centering
\includegraphics[width=0.95\textwidth]{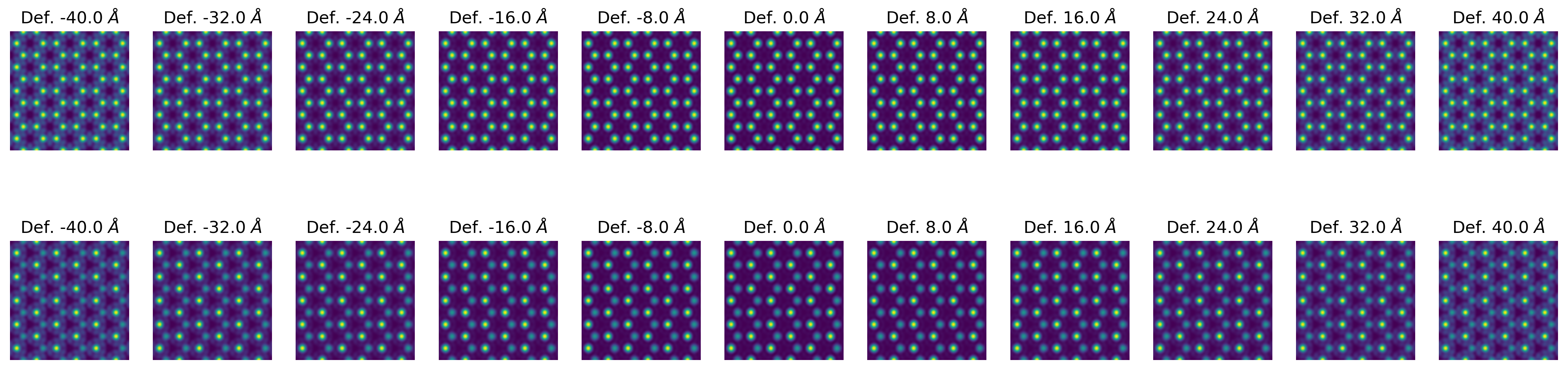}
\caption{Defocus series for monolayer (top row) and AA$'$ stacked bilayer (bottom row) h-BN. Defocus values range from $-40$ to $+40$ \AA{}. The monolayer shows strong B vs N contrast that varies with defocus, while the bilayer shows uniform intensity across both column types.}
\label{fig:defocus}
\end{figure}

Figure~\ref{fig:defocus} shows simulated HAADF images for both monolayer and bilayer h-BN across a range of defocus values. The bottom row shows single layer h-BN, where the intensity difference between boron and nitrogen atoms is clearly visible and varies significantly with defocus. In focus, the contrast between B and N is maximized, while over and underfocus show a drop in intensity.

The top row shows AA$'$ stacked bilayer h-BN. Notably, the two types of atomic columns (B-N and N-B) show essentially identical intensity across the entire defocus range. This indicates that the order of atoms along the beam direction does not significantly affect the HAADF intensity for this stacking configuration.

\subsection{Quantitative Analysis: Line Profiles}

\begin{figure}[h]
\centering
\includegraphics[width=0.9\textwidth]{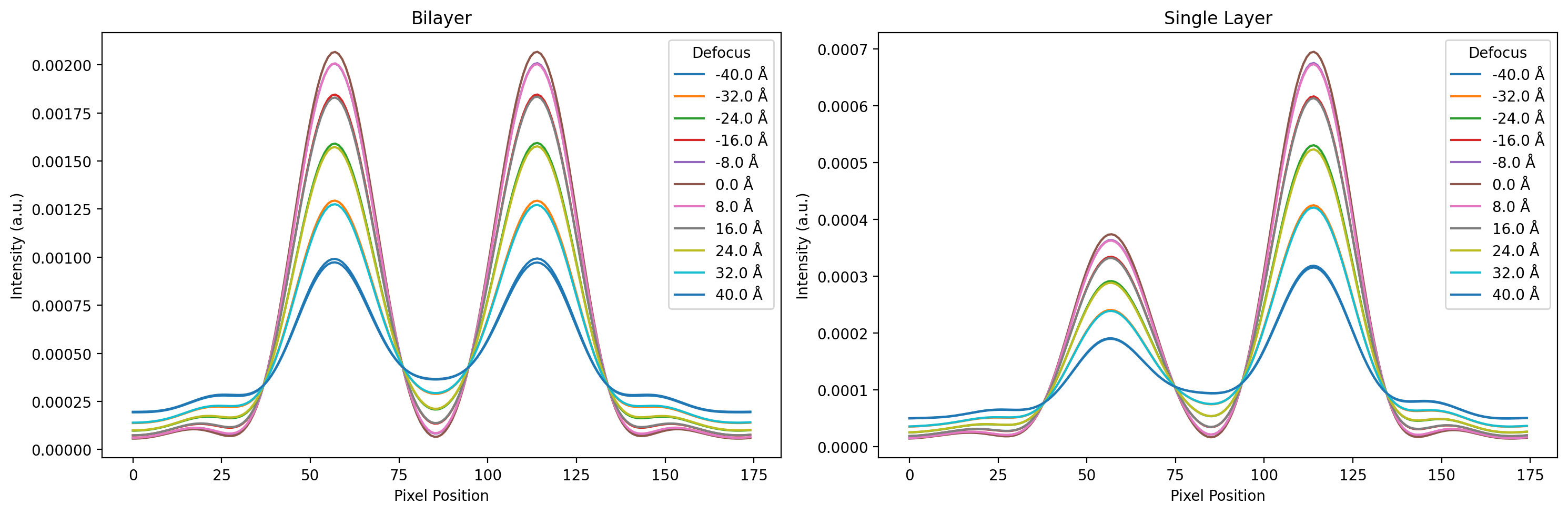}
\caption{Line profiles across atomic columns for different defocus values. Left: AA$'$ bilayer h-BN showing uniform peak heights for both column types. Right: Monolayer h-BN showing contrast difference between B (low) and N (high) peaks. }
\label{fig:profiles}
\end{figure}

To illustrate these intensity changes more quantitatively, line profiles were extracted perpendicular to the atomic columns for each defocus value. Figure~\ref{fig:profiles} shows these line profiles for both monolayer and bilayer structures.

For the monolayer, the alternating B and N peaks are clearly distinguishable, with nitrogen showing higher intensity. The magnitude of this difference varies with defocus, as expected from the defocus series, but the intensity difference remains significant for all defocus values.

For the bilayer, the two column types show nearly identical peak heights across all defocus values. Any residual differences are much smaller than the B-N difference in the monolayer and are comparable to the noise level in the simulations.

\subsection{Peak Height Difference Analysis}

\begin{figure}[h]
\centering
\includegraphics[width=0.6\textwidth]{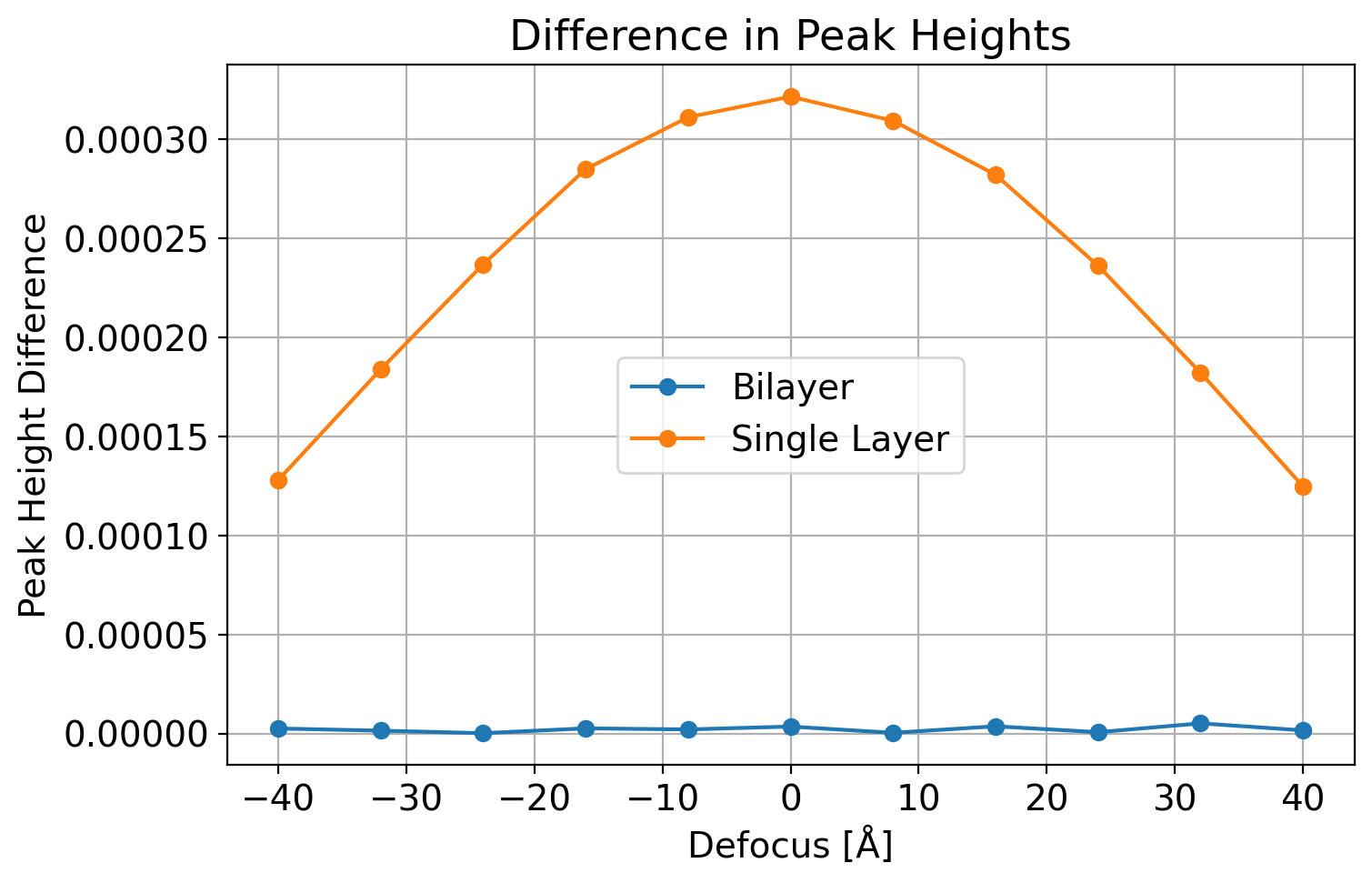}
\caption{Peak height difference as a function of defocus. The monolayer shows systematic variation with defocus (orange), while the bilayer shows no significant trend (blue), with differences comparable to noise level.}
\label{fig:peak_diff}
\end{figure}

To further quantify the column intensity differences, we extracted the peak heights from the line profiles and calculated the difference between the two peaks. Figure~\ref{fig:peak_diff} plots this peak height difference as a function of defocus for both monolayer and bilayer structures.

The monolayer shows a clear trend: the B-N intensity difference varies systematically with defocus, reaching a maximum at focus and decreasing in either direction. In contrast, the bilayer shows essentially no systematic variation with defocus. The peak height differences are small (on the order of the simulation noise) and do not follow any clear trend. This confirms that the B-N and N-B columns in AA$'$ stacked bilayer h-BN produce indistinguishable intensities in HAADF imaging.

\subsection{Simplified Comparison at Optimal Focus}

\begin{figure}[h]
\centering
\includegraphics[width=0.6\textwidth]{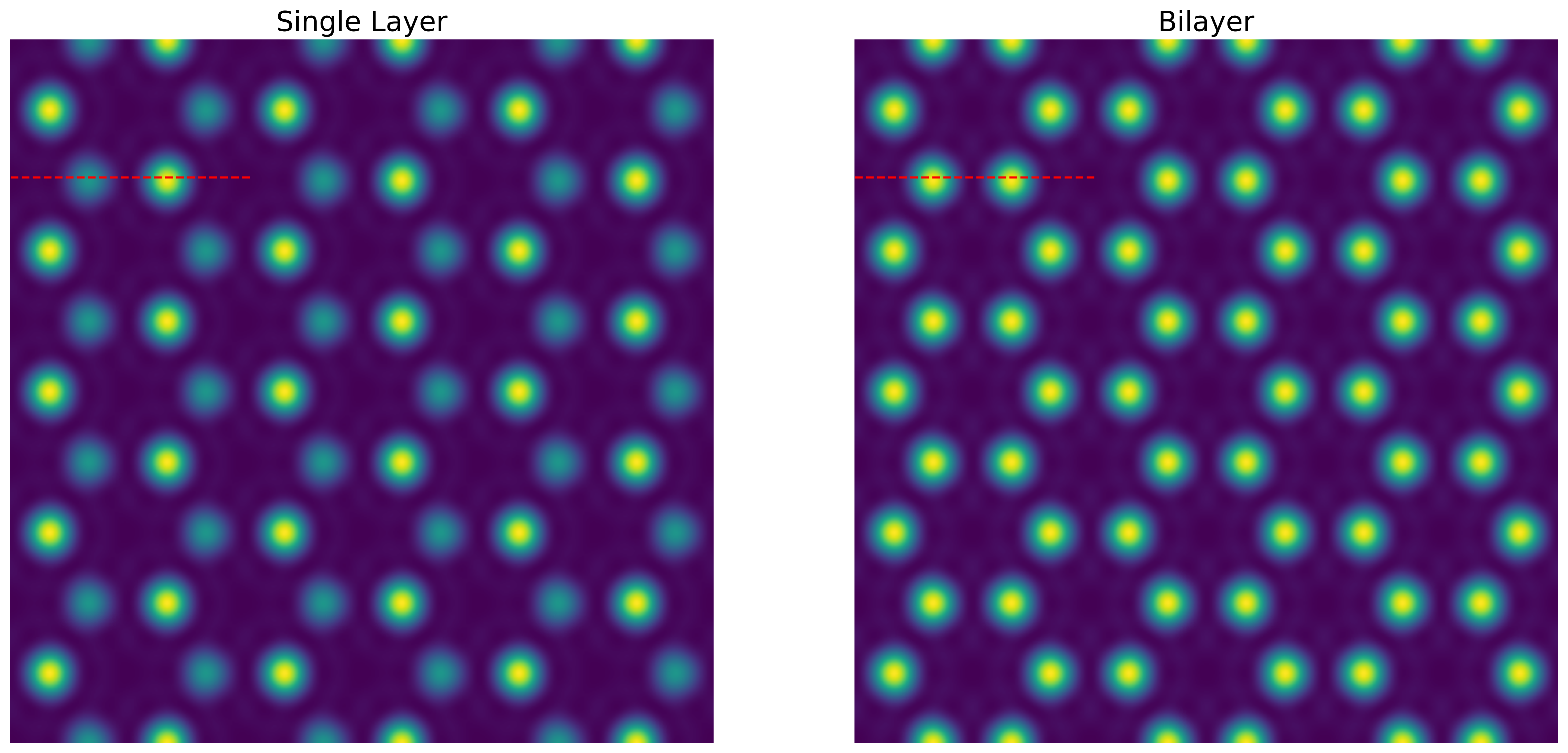}
\caption{Direct comparison of simulated HAADF images at optimal focus. Left: Monolayer h-BN showing distinct B and N intensities. Right: AA$'$ bilayer h-BN showing uniform intensity for both column types.}
\label{fig:comparison}
\end{figure}

For clarity, Figure~\ref{fig:comparison} presents a direct comparison of monolayer and bilayer h-BN at a single defocus value (in focus). The monolayer clearly shows two distinct intensity levels corresponding to B and N atoms. The bilayer shows uniform intensity across both column types, confirming that AA$'$ stacking produces no distinguishable intensity difference between B-N and N-B columns.

\begin{figure}[h]
\centering
\includegraphics[width=0.6\textwidth]{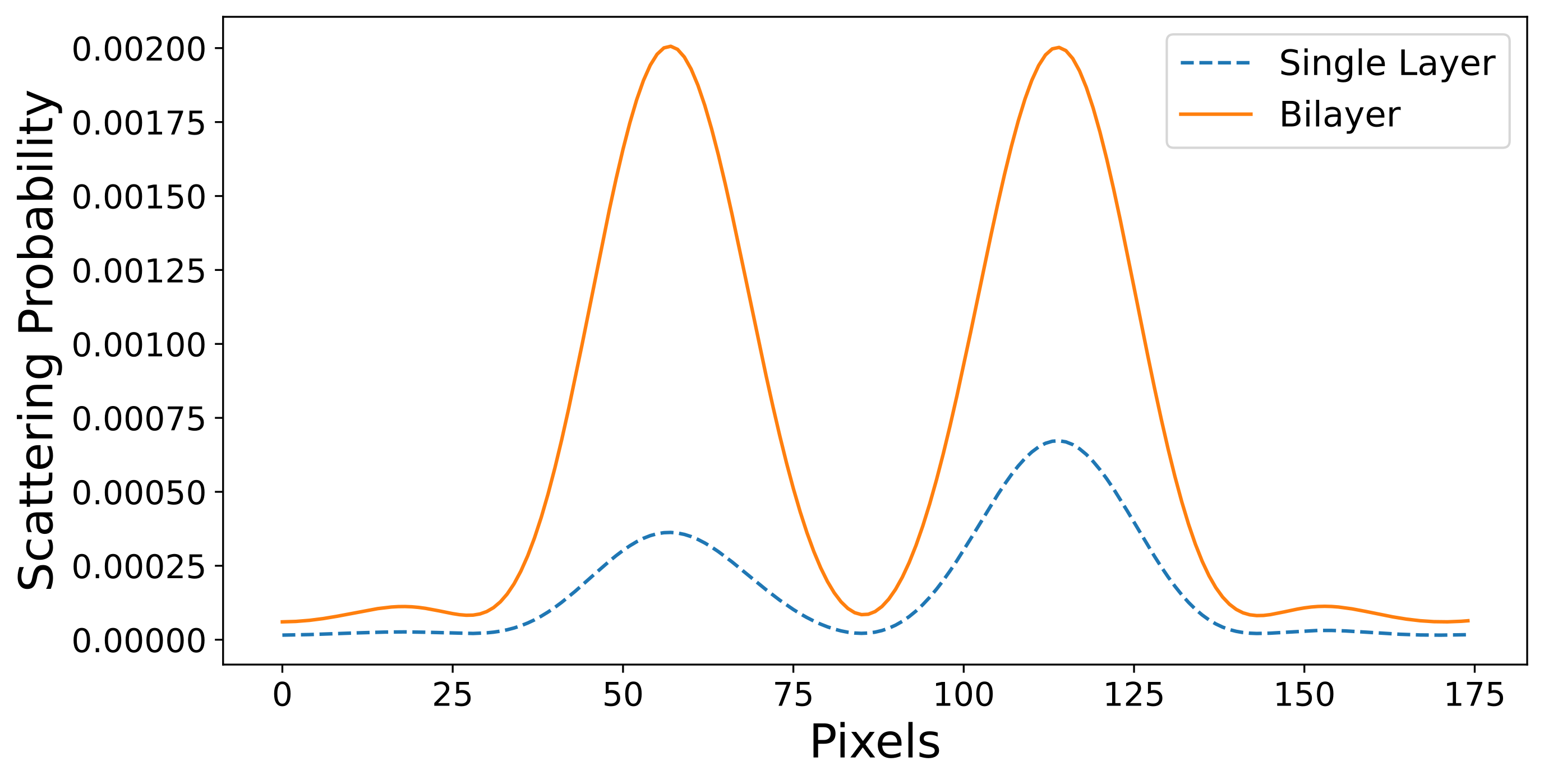}
\caption{Line profiles at optimal focus. The monolayer (blue) shows clear alternating B and N peaks, while the bilayer (orange) shows uniform peak heights.}
\label{fig:profile_comparison}
\end{figure}

Figure~\ref{fig:profile_comparison} shows the corresponding line profiles, making the intensity uniformity in the bilayer even more apparent.

\subsection{Convergence Testing}

\begin{figure}[h]
\centering
\includegraphics[width=0.6\textwidth]{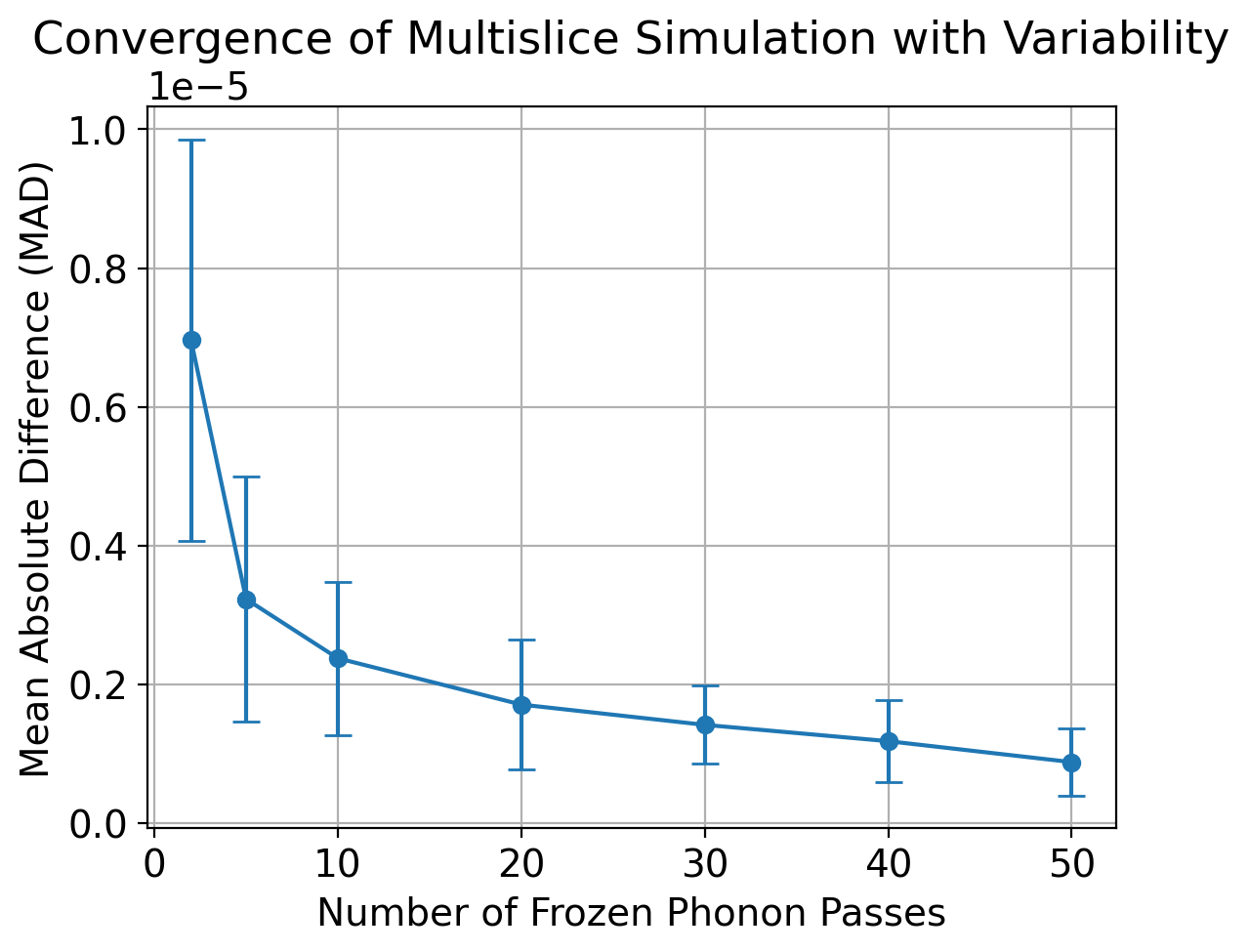}
\caption{Convergence test showing mean absolute difference (with standard deviation) between 15 independent simulation runs as a function of frozen phonon configurations. Convergence is achieved by 20--30 configurations, well below the 40 used in the main simulations.}
\label{fig:convergence}
\end{figure}

To validate that the observed intensity uniformity in the bilayer is not an artifact of insufficient sampling, we performed convergence testing by varying the number of frozen phonon configurations. Figure~\ref{fig:convergence} shows the mean absolute difference and standard deviation between 15 independent simulation runs as a function of the number of frozen phonon modes.

The results show that convergence is achieved with approximately 20--30 frozen phonon configurations. At 40 configurations (used in the main simulations), the variation between runs is much smaller than the intensity differences we are analyzing. This confirms that the simulations are adequately converged and that the observed intensity uniformity in AA$'$ bilayer h-BN is a real physical result, not a computational artifact.

\subsection{Conclusions}

These multislice simulations establish several key findings. First, AA$'$ stacked bilayer h-BN shows uniform intensity across both column types (B-N and N-B), indicating that the order of atoms along the beam direction does not significantly affect HAADF intensity for this stacking configuration. Second, this behavior is independent of defocus: unlike the monolayer, where B-N contrast varies systematically with defocus, the bilayer shows no such variation. Third, the simulations are well-converged, with 40 frozen phonon configurations producing simulation uncertainty much smaller than the intensity differences being analyzed. Finally, the Z$^2$ approximation holds for AA$'$ stacking, with intensity determined by the total atomic number content of the column rather than the specific ordering of atoms.

These results provide the foundation for interpreting experimental images of twisted bilayer h-BN. The uniform intensity of AA$'$ stacking serves as a baseline for identifying other stacking configurations (AA, AB Nitrogen, AB Boron) that do show distinct intensity signatures, as discussed in the main text.

\subsection{Code Availability}

The simulation code and structure files are available at: \href{https://github.com/ondrejdyck/atomic-precise-hbn-sculpt}{github.com/ondrejdyck/atomic-precise-hbn-sculpt}

\end{document}